\documentclass{aa}  

\usepackage{graphicx}
\usepackage{natbib}
\usepackage{rotating}
\usepackage[toc,page]{appendix}
\usepackage{multirow}
\usepackage{placeins}
\usepackage{epstopdf}
\usepackage{epsfig}
\usepackage{amssymb}
\usepackage{amsmath}
\usepackage{supertabular}
\usepackage{longtable}
\usepackage{ccaption}
\def\onecolumn{%
  \global\columnwidth\textwidth
  \global\hsize\columnwidth
  \global\linewidth\columnwidth
  }

\newcommand{\gTM}{\ensuremath{\gamma_{TM}}}

\newcommand{\gLM}{\ensuremath{\gamma_{LM}}}

\newcommand{\ALT}{\ensuremath{A_{LT}}}
\newcommand{\BLT}{\ensuremath{B_{LT}}}

\newcommand{\gLT}{\ensuremath{\gamma_{LT}}}
\newcommand{\intLT}{\ensuremath{\sigma_{LT}}}

\newcommand{\lxxl}{\ensuremath{L_{XXL}}}

\newcommand{\lbol}{\ensuremath{L_{bol}}}

\newcommand{\Th}{\ensuremath{\hat{T}}}
\newcommand{\Lh}{\ensuremath{\hat{L}}}
\newcommand{\lik}{\ensuremath{\cal {L}}}
\newcommand{\XMM}{{\em XMM}}

\usepackage{txfonts}
\usepackage{graphicx}
\usepackage{natbib}
\usepackage{rotating}
\usepackage[toc,page]{appendix}
\usepackage{caption}
\usepackage{float}
\floatstyle{plaintop}
\restylefloat{table}
\restylefloat{table}
\begin{document}

   \title{The XXL Survey\thanks{Based on observations obtained with
       {\em XMM-Newton}, an ESA science mission with instruments and
       contributions directly funded by ESA Member States and NASA.}}

   \subtitle{III. Luminosity-temperature relation of the Bright Cluster Sample}

   \author{P. A. Giles,
          \inst{1}
          B. J. Maughan,
          \inst{1}
          F. Pacaud,
          \inst{2}
          M. Lieu,
          \inst{3}
          N. Clerc,
          \inst{4}
          M. Pierre,
          \inst{5}
          C. Adami,
          \inst{6}
          L. Chiappetti,
          \inst{7}
          J. D\'emocl\'es,
          \inst{3}
          S. Ettori,
          \inst{8}
          \inst{9}
          J. P. Le F\'evre,
          \inst{5}
          T. Ponman,
          \inst{3}
          T. Sadibekova,
          \inst{6}
          G. P. Smith,
          \inst{3}
          J. P. Willis,
          \inst{10}
          F. Ziparo
          \inst{3}
        }

   \institute{School of Physics, HH Wills Physics Laboratory, Tyndall Avenue, Bristol, BS8 1TL,
  UK
              \\
             \email{P.Giles@bristol.ac.uk}
             \and
             Argelander-Institut fur Astronomie, University of Bonn,
             Auf dem Hugel 71, D-53121 Bonn, Germany \and
             School of Physics and Astronomy, University of
             Birmingham, Edgbaston, Birmingham, B15 2TT, UK \and
             Max Planck Institut fur Extraterrestrische Physik,
             Postfach 1312, D-85741 Garching bei Munchen, Germany \and
             Laboratoire AIM, CEA/DSM/IRFU/Sap, CEA Saclay, 91191
             Gif-sur-Yvette, France \and 
             LAM (Laboratoire d'Astrophysique de Marseille) UMR 7326m
             Aix-Marseille Universite, CNRS, F-13388 Marseille, France
             \and
             INAF, IASF Milano, via Bassini 15, I-20133 Milano, Italy
             \and 
             INAF, Osservatorio Astronomico di Bologna, via Ranzani 1,
             1-40127 Bologna, Italy \and
             INFN, Sezione di Bologna, viale Berti Pichat 612, 1-40127
             Bologna, Italy \and
             Department of Physics and Astronomy, University of
             Victoria, 3800 Finnerty Road, Victoria, BC, Canada \\
             }

   \date{\today}

% \abstract{}{}{}{}{} 
% 5 {} token are mandatory
 
  \abstract
   {The {\em XXL} Survey is the largest homogeneous survey carried out
     with {\em XMM-Newton}.  Covering an area of 50 deg$^{2}$, the survey
     contains several hundred  galaxy clusters out to a redshift of
     $\approx$2 above an X-ray flux limit of $\sim$5$\times$10$^{-15}$
     erg cm$^{-2}$ s$^{-1}$.  This paper belongs to the first series
     of XXL papers focusing on the bright cluster sample.}
   {We  investigate the luminosity-temperature ($LT$)
     relation for the brightest clusters detected in the XXL Survey,
     taking fully into account the selection biases.  We
     investigate the form of the $LT$ relation, placing constraints on
     its evolution.}
   {We have classified the 100 brightest clusters in the XXL
     Survey based on their measured X-ray flux.  These 100 clusters
     have been analysed to determine their luminosity and temperature to
     evaluate the $LT$ relation.  We  used three methods to fit  the
   form of the $LT$ relation, with two of these methods providing a
   prescription to fully take into account the selection effects of
   the survey.  We measure the evolution of the $LT$ relation
   internally using the broad redshift range of the sample.}
   {Taking fully into account selection effects, we find a slope of
     the bolometric $LT$ relation of B$_{\rm LT}$=3.08$\pm$0.15,
     steeper than the self-similar expectation (B$_{\rm LT}$=2).  Our
     best-fit result for the evolution factor is
     $E(z)^{1.64\pm0.77}$, fully consistent with ``strong
     self-similar'' evolution where clusters scale self-similarly with
     both mass and redshift.  However, this result is marginally
     stronger than ``weak self-similar'' evolution, where clusters
     scale with  redshift alone.  We investigate the sensitivity of our
     results to the assumptions made in our fitting model, finding
     that using an external $LT$ relation as a low-z baseline can have
     a profound effect on the measured evolution.  However, more
     clusters are needed in order to break the degeneracy between the
     choice of likelihood model and mass-temperature relation on the
     derived evolution.} 
   {}

   \keywords{}
   \authorrunning{P. Giles et al.}
   \titlerunning{XXL III: $LT$ Relation of the Bright Cluster Sample}
   \maketitle
   
%
%________________________________________________________________

\section{Introduction}
\label{sec:intro}

Under the assumption of self-similarity \citep{1986MNRAS.222..323K},
simple scaling laws can be derived between various properties of
galaxy clusters.  These scaling laws are advantageous as they provide a
cheap way of measuring the masses of large samples of clusters, an
important ingredient for cosmological studies using galaxy clusters
\citep[see][for a review]{2011ARA&A..49..409A}.  
One of the most explored scaling relations is that between the X-ray
luminosity (L) and temperature (T), expected to follow a
relationship of $L\propto T^{2}$.  However, a multitude of
studies have found that the slope of the $LT$ relation is
$\sim$3  \citep[e.g.][]{2009A&A...498..361P,2011A&A...535A.105E,2013A&A...558A..75T,con14},
steeper than the self-similar expectation.  One of the main
explanations for the deviation from this theoretical expectation is
energy input by non-gravitational processes during cluster formation at early
times, including pre-heating, supernovae feedback, and heating from
active galactic nuclei (AGN) at high redshift.  These
non-gravitational processes should have the strongest effect in lower
mass systems owing to their shallower potential wells that expel gas
from the inner regions and suppress the luminosity.  Observations
have shown further steepening of the $LT$ relation at the low-mass
regime \citep[e.g.][]{2004MNRAS.350.1511O,2009ApJ...693.1142S}.
However, recent work, using methods to correct for sample selection
effects, has suggested that the slope of the $LT$
relation on group scales is consistent with massive clusters
\citep[e.g.][]{2015A&A...573A..75B,2015A&A...573A.118L}. 

X-ray flux limited samples suffer from two forms of selection bias,
Malmquist bias, where higher luminosity clusters are preferentially
selected out to higher redshifts, and Eddington bias, where in the
presence of intrinsic or statistical scatter in luminosity for a given
mass, objects above a flux limit will have above average luminosities
for their mass. This effect is amplified by the steep slope of the
cluster mass function, which results in a net movement of lower mass
objects into a flux limited sample.  The net effect on the $LT$
relation is to bias the normalisation high  and the slope low
\citep[see][for a review]{2011ARA&A..49..409A}.  Therefore,
taking these biases into account is paramount when modelling cluster
scaling relations in order to uncover the true nature of any
non-gravitational heating driving departures from self-similar
behaviour  with mass or redshift.  Although scaling relation
studies have had a rich history, only a relatively small number of
published relations attempt to account for selection biases
\citep[e.g.][]{2006ApJ...648..956S,2007MNRAS.382.1289P,2009A&A...498..361P,2009ApJ...692.1033V,2012A&A...547A.117A,2015A&A...573A..75B,2015A&A...573A.118L},
while \cite{2010MNRAS.406.1773M} provides the most robust handling of
selection effects to date.            

In the self-similar model, with cluster properties measured within
overdensity radii defined relative to the critical density, the
evolution of the scaling relations can be parameterised by the factor
$E(z)$ (where $E(z)=\sqrt{\Omega_{\rm M}(1 + z)^3 + (1 - \Omega_{\rm
    M} - \Omega_{\Lambda})(1 + z)^2 + \Omega_{\Lambda}}$).  In this
framework, the evolution of the normalisation of the $LT$
relation goes as $E(z)^{\gamma_{LT}}$, where $\gamma_{LT}\equiv1$ if
clusters scale self-similarly with both mass and redshift, or
$\gamma_{LT}\approx 0.42$ if the observed mass dependence of the
cluster baryon fraction is also included
\citep{2014MNRAS.437.1171M}.  These two reference values for 
$\gamma_{LT}$ assume the same underlying evolution driven by the
changing critical density of the Universe; the difference arises
purely from the algebraic combination of the scaling laws of gas mass,
temperature, and gas structure with total mass that are used to
construct the $LT$ relation \citep{2014MNRAS.437.1171M}.

We refer to $\gamma_{LT}=1$ as ``strong self-similar'' evolution,
reflecting the fact that it is based on the assumption that clusters
scale self-similarly with both mass and redshift, while
$\gamma_{LT}=0.42$ is referred to as ``weak self-similar'' evolution
and  assumes clusters only scale self-similarly with redshift. The
latter is a more realistic prediction of the self-similar evolution
against which to test for departures in the observed evolution of
clusters in the $LT$ plane.

Understanding how scaling relations evolve with redshift is important
for two main reasons: (i) scaling relations must be well-calibrated
at high redshift to provide mass estimates for cosmology and (ii) the relations
provide insight into the history of heating mechanisms in clusters.
A consensus on the evolution from observations has yet to be
achieved; various studies find an evolution consistent with
self-similar  
\citep[e.g.][]{2002ApJ...578L.107V,2006MNRAS.365..509M,2007MNRAS.382.1289P},
while others find departures from the self-similar expectation 
\citep[e.g.][]{2004A&A...417...13E,2005ApJ...633..781K,2011A&A...535A...4R,2012MNRAS.424.2086H,2014MNRAS.444.2723C}.
Again, the importance of taking into account selection effects is crucial
for understanding evolution and recent work has shown that departures from
self-similar evolution can be explained by selection biases
\citep{2007MNRAS.382.1289P,2010MNRAS.406.1773M,2012MNRAS.421.1583M}.
Although these processes need to be understood to explain the
differences between the observed and theoretical prediction, further
considerations must be made to take into account the selection effects. 

In order to address the points raised above, we have carried out the
largest X-ray survey undertaken by \XMM-Newton.  The XXL Survey
\citep[][hereafter Paper I]{test1}, is a 50 deg$^2$ survey with
an X-ray sensitivity of $\sim$5$\times$10$^{-15}$ erg s$^{-1}$
cm$^{-2}$ (for point-like sources), in the [0.5-2] keV band.  With the
aim of detecting several hundred clusters out to a redshift of
$\approx$2, this survey provides a unique opportunity to constrain
cluster scaling relations that fully accounts for selection effects, and
in the future will place robust constraints on cosmological parameters
(see Paper I).     

In this paper, part of the first release of XXL results, we
investigate the form of the luminosity-temperature
relation ($LT$) for a sample of the 100 brightest clusters
detected in the XXL Survey.  The $LT$ relation will be derived taking
 the selection function of the cluster sample fully into account.  The
outline of this paper is as follows.  In $\S$~\ref{sec:sample} we
discuss the data preparation and sample selection.
Section~\ref{sec:analysis} outlines the cluster analysis.  In
$\S$~\ref{sec:results} we present our results and derive the sample
and bias-corrected scaling relations.  Our discussion and conclusions
are presented in $\S$~\ref{sec:disc} and $\S$~\ref{sec:conc},
respectively.  Throughout this paper we assume a WMAP9
cosmology of $\Omega_{\rm M}$=0.282, $\Omega_{\Lambda}$=0.719, and H$_{0}$=69.7 
\citep{2013ApJS..208...19H}.

%__________________________________________________________________

\section{Data processing and sample selection}
\label{sec:sample}

The data processing and sample selection are fully detailed in
\citet[][hereafter Paper II]{test2}, and is briefly summarised here.
The XXL Survey contains 542 {\em XMM} pointings covering 50.8 deg$^2$.
After light-curve filtering \citep[following][]{2002A&A...394..375P},
rejection of bad pointings and exclusion of pointings with high
background levels, the total XXL area spans 46.6 deg$^{2}$ (from 454
pointings).  Images, exposure maps, and detector masks were generated
and processed using the {\sc Xamin} pipeline
\citep[][]{2006MNRAS.372..578P,2012MNRAS.423.3561C}.  {\sc SExtractor} 
was then run to generate a conservative source catalogue and source
masks, followed by a dedicated {\em XMM} maximum likelihood fitting
procedure to determine likelihood ratios to assess the detection and
source extent probabilities.  Extended sources were defined with an
extent larger than 5$^{\prime\prime}$ and extension likelihood larger
than 15.  These extended sources were then separated into two classes,
the C1 class with extension likelihood larger than 33 and detection
likelihood larger than 32, and the C2 class with an extension
likelihood 15 $<$ EXT\_LH $<$ 32.  

The sample used in this work is the 100 brightest clusters
(hereafter XXL-100-GC\footnote{A master XXL-100-GC cluster catalogue
  will also be available in electronic form at
  \url{http://cosmosdb.isaf-milano.inaf.it/XXL/} and via the {\em XMM}
  XXL DataBase at \url{http://xmm-lss.in2p3.fr}})
selected from the  source list generated above.
Count rates were estimated from a growth curve analysis
\citep[GCA, following][]{2012MNRAS.423.3561C} within a 60$^{\prime\prime}$
aperture, and converted into fluxes using an energy conversion
factor (assuming $T$=3 keV, z=0.3, and Z=0.3) of 9.04$\times$10$^{-13}$,
or 1.11$\times$10$^{-13}$ for clusters falling on the damaged mos1
chip. The clusters were ranked in order
of decreasing flux and the brightest 100 clusters were selected.  We
note that five clusters within the XXL-100-GC fell on observations with
high periods of flaring.  The five flared observations were not used
and instead the next five brightest objects were included.  The
clusters \texttt{XLSSC} 113, 114, 115, 550, and 551 were replaced with
\texttt{XLSSC} 091, 506, 516, 545, and 548.  Taking this into account, our
selection corresponds to a flux limit of F$_{\rm
  X,cut}$=3$\times$10$^{-14}$ ergs s$^{-1}$ cm$^{-2}$. The final
XXL-100-GC sample consists of 96 C1 and 4 C2 clusters,
respectively, with 51 falling within the northern field of the XXL
footprint, and 49 within the southern field.  Following a robust redshift
validation (see Paper I), the XXL-100-GC span a redshift range of
0.04$\leq$z$\leq$1.05, with 98 spectroscopic and 2 photometric
redshifts.  The sample and the derived properties can be
found in Table~\ref{tab:results}. 

%                                     Two column figure (place early!)
%______________________________________________ Gamma_1 (lg rho, lg e)

\section{Analysis}
\label{sec:analysis}

In this section we describe the cluster analysis process used in this
work.  

The extent of the cluster emission was defined as the radius beyond
which no significant cluster emission is detected using a threshold of 
0.5$\sigma$ above the background level.  This is intended to provide a
conservative estimate of the radius beyond which there is no
significant cluster emission.  A one-dimensional (1D)
surface brightness profile of each camera was produced.  A
background annulus was defined with an inner radius of
250$^{\prime\prime}$ and modelled by a flat (particle) and vignetted
(X-ray) component.  The two model components were then fit to the
radial profile in the background region using the $\chi^{2}$
statistic and then summed to produce a total background model.  The radial
bins were constructed to be 15.4$^{\prime\prime}$ in width, chosen to ensure
greater than 20 counts per bin and with enough bins to perform the
fitting procedure.  The radial profile for each camera was then summed
to produce an overall profile.  An initial source extent was
determined based on the 0.5$\sigma$ radius, and the background
fitting repeated using the extent as the inner radius for the
background annulus.  This process was iterated until the source
extent radius changed by less than 1\%.

\begin{figure}[t]
\begin{center}
\includegraphics[width=10cm, clip=true, trim={0.5cm 0 0 5.5cm}]{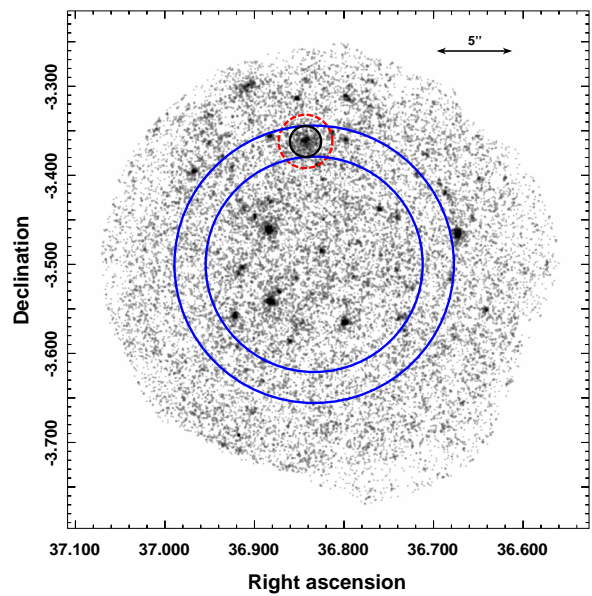}
\end{center}
\vspace{-3cm}
\caption[]{\small{Example of the background method used in the
    spectral analysis (see $\S$~\ref{sec:analysis}).  The image of the
    cluster \texttt{XLSSC} 010 with the regions showing the spectral extraction
    region (black circle), cluster detection radius (dashed red
    circle), and the background annulus centred on the aimpoint of the
    observation (blue annulus).}\label{fig:bgexample}}
\end{figure}

To account for the background in the spectral analysis, local
backgrounds were used.  However, owing to the survey detection of the
clusters, they are detected at a range of off-axis positions on the
{\em XMM} cameras.  Therefore, if possible, an annulus centred on the
aimpoint of the observation (BG$_{\rm aim}$) with a width equal to
the diameter of the spectral extraction region was used.  This ensures
that the local background is taken at the same off-axis position as
the cluster, which helps to reduce the systematic uncertainties due to the
radial dependencies of the background components.  To ensure that no
cluster emission was included in the background subtraction, a region
centred  on the cluster with radius equal to the cluster extent (see
above) was excluded.  Figure~\ref{fig:bgexample} shows an example of
this local background for one of our clusters.  However,
if this method was not possible (owing to close proximity to the
aimpoint or a large cluster extent), an annulus centred on the
cluster (BG$_{\rm clust}$) with inner radius  equal to the cluster
extent and an outer radius of 400$^{\prime\prime}$ was used for the
local background.  The clusters \texttt{XLSSC} 060 and \texttt{XLSSC}
091 had a source extent larger than 400; therefore, an outer radius
of 500$^{\prime\prime}$ and 800$^{\prime\prime}$, respectively, was
used for the cluster centred local background.

Cluster spectra were extracted for each of the {\em XMM} cameras and fits
were performed in the 0.4-7.0 keV band with an absorbed APEC
\citep{2001ApJ...556L..91S} model (v2.0.2) with the absorbing column
fixed at the Galactic value \citep{2005A&A...440..775K}. The spectra
for each camera were fit simultaneously with the temperature of the
APEC components tied together.  The fits were performed using {\tt XSPEC}
(v12.8.1i) and the abundance table from \cite{1989GeCoA..53..197A}.
Because of the low number of counts for many of the clusters, the spectra
were fitted using the cstat statistic.  The background spectra were
grouped such that they contained at least 5 counts per bin, and this
grouping applied to the source spectra.  A similar method was employed
and justified in \cite{2005MNRAS.363..675W} who analysed a sample of
12 galaxy groups and clusters in the XMM-LSS.  Throughout the spectral
analysis we assumed a fixed metal abundance of Z=0.3Z$_{\odot}$.  

The cluster temperatures are derived within 300kpc for each cluster,
denoted as $T_{\rm 300kpc}$.  This radius represents the largest
radius for which a temperature could be derived for the entire cluster
sample.  The normalisations for each camera were free in
the spectral fit,  with the pn camera used to calculate the
luminosity.  We denote the luminosity within 300kpc as $L^{\rm
  XXL}_{300kpc}$, where the superscript XXL refers to the [0.5-2.0]
keV band (cluster rest frame).  Luminosities quoted within
$r_{500}$ are extrapolated from 300kpc out to $r_{\rm 500}$ by
integrating under a $\beta$-profile assuming
$r_{\rm c}$=0.15$r_{500}$ and $\beta$=0.667.  The $\beta$-profile
parameters were chosen to match those used in Paper II.  We note that the
uncertainties on the luminosity are scaled by this extrapolation, but
do not include any uncertainty on the $\beta$-profile parameters.  The
impact of these assumptions is tested in section~\ref{sec:extrap}.
Values for cluster r500 are calculated using a mass-temperature
relation ($r_{\rm 500,MT}$, see below).  We denote $L^{\rm XXL}_{500,MT}$ as
the luminosity in the [0.5-2.0] keV band (cluster rest frame) within
$r_{\rm 500,MT}$, and $L^{\rm bol}_{500,MT}$ as the bolometric
luminosity within $r_{\rm 500,MT}$.  All the properties quoted include
the cluster core because the exclusion of the cluster core is not
possible for all 100 clusters.  The results of our spectral analysis
are given in Table~\ref{tab:results}.   The cluster
\texttt{XLSSC} 504 was dropped because of the unconstrained errors
reported by {\tt XSPEC}.   

A mass-temperature ($M_{WL}-T$) relation is used in this work for two
purposes: (i) to calculate $r_{500,MT}$ for the XXL-100-GC and (ii) to convert
 the mass function to a temperature function (see
$\S$~\ref{sec-1}) for the bias correction.  The $M_{WL}-T$ relation is
presented in \citet[][hereafter Paper IV]{test4}, based upon weak
lensing cluster masses and the temperatures presented in this work.
Briefly, the $M_{WL}-T$ relation was determined using
37 XXL-N clusters that fell within the Canada-France-Hawaii Telescope
Lensing Survey (CFHTLenS) footprint, utilising the CFHTLenS shear
catalogue \citep{2012MNRAS.427..146H,2013MNRAS.433.2545E} for the mass
measurements.  In order to increase the statistics and the mass range
covered, the XXL-100-GC is combined with 10 groups from the COSMOS
survey \citep{2013ApJ...778...74K}, and 50 massive clusters from the
Canadian Cluster Comparison Project
\citep[CCCP,][]{2015MNRAS.449..685H}.    
     
The combined $M_{WL}-T$ relation is fitted with a power-law of the
form (assuming self-similar evolution)
\begin{align} 
\hspace{1.9cm}log_{10}(M_{WL}E(z)) & = A_{MT} + B_{MT}log_{10}(T)
\end{align}
using the Bayesian code {\em linmix\_err}
\citep{2007ApJ...665.1489K}.  In Paper IV we find $A_{MT}$=13.56$^{+0.10}_{-0.08}$
and $B_{MT}$=1.69$^{+0.12}_{-0.13}$.  See Paper IV for further
discussions on the results from the $M_{WL}-T$ relation.

\section{Results}
\label{sec:results}

The results are presented in the form of a study of the $LT$ relation,
taking the selection effects fully into account.  We present two
implementations to account for the selection effects:  (i)
an updated method of the \cite{2007MNRAS.382.1289P} implementation,
defined as the XXL likelihood, and  (ii) a method based upon
\cite{2010MNRAS.406.1759M}, defined as the M10 likelihood.
We first present the sample $LT$ relation, not correcting for the
selection effects.  This allows for comparison of the $LT$ relation
when taking into account selection effects,  what impact this
has on the derived scaling relation, and also for the comparison with
previously published relations that do not account for selection
effects. 

For the analysis of the scaling relations, we account for the fact
that the likelihood curve for a measured temperature is approximately
Gaussian in log space (consistent with the asymmetric errors usually
found in temperature measurements).  We use the method of
\cite{2012A&A...546A...6A} to convert the generally asymmetric errors
reported by {\tt XSPEC} into a log-normal likelihood.  We note that the errors
on the temperature given in Table~\ref{tab:results} are those reported
by {\tt XSPEC}.   

\begin{table*}
\begin{center}
\caption[]{\small{X-ray properties of the XXL-100-GC.  $r_{\rm 500,MT}$ is
    estimated from the $M_{WL}-T$ relation in Sect.~\ref{sec:analysis}.
    $L^{\rm XXL}_{500,MT}$ and $L^{\rm bol}_{500,MT}$ are extrapolated out to
    $r_{500,MT}$ by integrating under a $\beta$-profile (see
    Sect.~\ref{sec:results}).  $\dagger$ indicates that only a
    photometric redshift was available for the cluster
    analysis.}\label{tab:results}}  
%\vspace{-0.3cm}
\begin{tabular}{lccccccc}
\hline\hline 
\texttt{XLSSC} Num & z & E(z) & $r_{500,MT}$ & $T_{\rm 300kpc}$ & $L^{\rm
  XXL}_{\rm 300kpc}$ & $L^{\rm XXL}_{500,MT}$ & $L^{\rm bol}_{500,MT}$ \\  
 & & & Mpc & (keV) & 10$^{43}$(ergs s$^{-1}$) & 10$^{43}$(ergs s$^{-1}$) & 10$^{43}$(ergs s$^{-1}$) \\
\hline
\texttt{XLSSC} 001 & 0.614 & 1.38 & 0.777 & 3.8$^{+0.5}_{-0.4}$ & 7.56$\pm$0.54  & 10.10$\pm$0.72 & 23.48$\pm$1.68 \\    
\texttt{XLSSC} 003 & 0.836 & 1.57 & 0.643 & 3.4$^{+1.0}_{-0.6}$ & 10.04$\pm$1.21 & 12.32$\pm$1.49 & 26.64$\pm$3.22 \\     
\texttt{XLSSC} 006 & 0.429 & 1.24 & 0.982 & 4.8$^{+0.5}_{-0.4}$ & 11.44$\pm$0.55 & 17.42$\pm$0.83 & 46.72$\pm$2.24 \\     
\texttt{XLSSC} 010 & 0.330 & 1.18 & 0.751 & 2.7$^{+0.5}_{-0.3}$ & 1.97$\pm$0.16  & 2.58$\pm$0.21  & 5.58$\pm$0.46 \\   
\texttt{XLSSC} 011 & 0.054 & 1.02 & 0.831 & 2.5$^{+0.5}_{-0.4}$ & 0.11$\pm$0.01  & 0.15$\pm$0.01  & 0.33$\pm$0.02 \\   
\texttt{XLSSC} 022 & 0.293 & 1.15 & 0.671 & 2.1$^{+0.1}_{-0.1}$ & 2.45$\pm$0.09  & 3.06$\pm$0.11  & 6.06$\pm$0.22 \\   
\texttt{XLSSC} 023 & 0.328 & 1.17 & 0.655 & 2.1$^{+0.3}_{-0.2}$ & 1.32$\pm$0.14  & 1.63$\pm$0.18  & 3.21$\pm$0.35 \\   
\texttt{XLSSC} 025 & 0.265 & 1.14 & 0.751 & 2.5$^{+0.2}_{-0.2}$ & 1.68$\pm$0.08  & 2.21$\pm$0.11  & 4.69$\pm$0.23 \\   
\texttt{XLSSC} 027 & 0.295 & 1.15 & 0.768 & 2.7$^{+0.4}_{-0.3}$ & 1.11$\pm$0.09  & 1.48$\pm$0.11  & 3.20$\pm$0.25 \\   
\texttt{XLSSC} 029 & 1.050 & 1.77 & 0.626 & 4.1$^{+1.0}_{-0.6}$ & 16.03$\pm$1.38 & 19.46$\pm$1.68 & 43.67$\pm$3.76 \\     
\texttt{XLSSC} 036 & 0.492 & 1.29 & 0.801 & 3.6$^{+0.5}_{-0.4}$ & 8.21$\pm$0.53  & 11.14$\pm$0.72 & 25.78$\pm$1.68 \\     
\texttt{XLSSC} 041 & 0.142 & 1.07 & 0.670 & 1.9$^{+0.1}_{-0.2}$ & 0.96$\pm$0.06  & 1.19$\pm$0.07  & 2.31$\pm$0.14 \\   
\texttt{XLSSC} 050 & 0.140 & 1.07 & 0.897 & 3.1$^{+0.2}_{-0.2}$ & 1.93$\pm$0.05  & 2.78$\pm$0.07  & 6.68$\pm$0.17 \\   
\texttt{XLSSC} 052 & 0.056 & 1.02 & 0.387 & 0.6$^{+0.0}_{-0.0}$ & 0.09$\pm$0.01  & 0.09$\pm$0.01  & 0.14$\pm$0.01 \\   
\texttt{XLSSC} 054 & 0.054 & 1.02 & 0.723 & 2.0$^{+0.2}_{-0.2}$ & 0.21$\pm$0.02  & 0.28$\pm$0.02  & 0.56$\pm$0.04 \\   
\texttt{XLSSC} 055 & 0.232 & 1.12 & 0.843 & 3.0$^{+0.3}_{-0.4}$ & 1.88$\pm$0.11  & 2.61$\pm$0.15  & 6.02$\pm$0.34 \\   
\texttt{XLSSC} 056 & 0.348 & 1.19 & 0.824 & 3.2$^{+0.5}_{-0.3}$ & 3.03$\pm$0.18  & 4.16$\pm$0.25  & 9.58$\pm$0.58 \\   
\texttt{XLSSC} 057 & 0.153 & 1.07 & 0.734 & 2.2$^{+0.3}_{-0.1}$ & 0.91$\pm$0.05  & 1.18$\pm$0.07  & 2.46$\pm$0.14 \\   
\texttt{XLSSC} 060 & 0.139 & 1.07 & 1.136 & 4.8$^{+0.2}_{-0.2}$ & 3.75$\pm$0.05  & 6.31$\pm$0.08  & 18.57$\pm$0.24 \\     
\texttt{XLSSC} 061 & 0.259 & 1.13 & 0.678 & 2.1$^{+0.5}_{-0.3}$ & 0.87$\pm$0.11  & 1.09$\pm$0.14  & 2.16$\pm$0.27 \\   
\texttt{XLSSC} 062 & 0.059 & 1.03 & 0.422 & 0.7$^{+0.1}_{-0.1}$ & 0.10$\pm$0.02  & 0.11$\pm$0.02  & 0.16$\pm$0.02 \\   
\texttt{XLSSC} 072 & 1.002 & 1.73 & 0.613 & 3.7$^{+1.1}_{-0.6}$ & 12.35$\pm$1.49 & 14.89$\pm$1.79 & 32.61$\pm$3.92 \\     
\texttt{XLSSC} 083 & 0.430 & 1.24 & 0.943 & 4.5$^{+1.1}_{-0.7}$ & 3.14$\pm$0.25  & 4.67$\pm$0.38  & 12.11$\pm$0.98 \\    
\texttt{XLSSC} 084 & 0.430 & 1.24 & 0.945 & 4.5$^{+1.6}_{-1.3}$ & 1.38$\pm$0.21  & 2.04$\pm$0.31  & 6.18$\pm$0.84 \\   
\texttt{XLSSC} 085 & 0.428 & 1.24 & 0.976 & 4.8$^{+2.0}_{-1.0}$ & 2.56$\pm$0.26  & 3.88$\pm$0.40  & 11.46$\pm$1.18 \\    
\texttt{XLSSC} 087 & 0.141 & 1.07 & 0.619 & 1.6$^{+0.1}_{-0.1}$ & 0.76$\pm$0.07  & 0.92$\pm$0.09  & 1.67$\pm$0.16 \\   
\texttt{XLSSC} 088 & 0.295 & 1.15 & 0.726 & 2.5$^{+0.6}_{-0.4}$ & 1.22$\pm$0.11  & 1.57$\pm$0.15  & 3.28$\pm$0.31 \\   
\texttt{XLSSC} 089 & 0.609 & 1.38 & 0.769 & 3.7$^{+1.6}_{-1.2}$ & 4.85$\pm$0.66  & 6.44$\pm$0.87  & 14.84$\pm$2.01 \\     
\texttt{XLSSC} 090 & 0.141 & 1.07 & 0.507 & 1.1$^{+0.1}_{-0.1}$ & 0.38$\pm$0.05  & 0.43$\pm$0.05  & 0.67$\pm$0.08 \\   
\texttt{XLSSC} 091 & 0.186 & 1.09 & 1.149 & 5.1$^{+0.2}_{-0.2}$ & 7.73$\pm$0.11  & 13.12$\pm$0.19 & 39.12$\pm$0.58 \\     
\texttt{XLSSC} 092 & 0.432 & 1.24 & 0.771 & 3.1$^{+0.8}_{-0.6}$ & 2.11$\pm$0.24  & 2.81$\pm$0.31  & 6.24$\pm$0.70 \\    
\texttt{XLSSC} 093 & 0.429 & 1.24 & 0.810 & 3.4$^{+0.6}_{-0.4}$ & 4.75$\pm$0.31  & 6.47$\pm$0.42  & 14.91$\pm$0.96 \\      
\texttt{XLSSC} 094 & 0.886 & 1.62 & 0.742 & 4.7$^{+1.3}_{-0.9}$ & 19.85$\pm$1.71 & 25.93$\pm$2.24 & 62.01$\pm$5.35 \\     
\texttt{XLSSC} 095 & 0.138 & 1.06 & 0.450 & 0.9$^{+0.1}_{-0.1}$ & 0.15$\pm$0.03  & 0.17$\pm$0.03  & 0.24$\pm$0.04 \\   
\texttt{XLSSC} 096 & 0.520 & 1.31 & 1.000 & 5.5$^{+2.0}_{-1.1}$ & 3.77$\pm$0.40  & 5.80$\pm$0.62  & 16.05$\pm$1.71 \\    
\texttt{XLSSC} 097 & 0.760 & 1.50 & 0.794 & 4.6$^{+1.5}_{-1.0}$ & 9.91$\pm$1.25  & 13.37$\pm$1.69 & 32.53$\pm$4.11 \\     
\texttt{XLSSC} 098 & 0.297 & 1.15 & 0.801 & 2.9$^{+1.0}_{-0.6}$ & 1.28$\pm$0.16  & 1.74$\pm$0.21  & 3.88$\pm$0.48 \\   
\texttt{XLSSC} 099 & 0.391 & 1.22 & 1.032 & 5.1$^{+3.1}_{-1.5}$ & 1.44$\pm$0.26  & 2.26$\pm$0.41  & 6.31$\pm$1.13 \\   
\texttt{XLSSC} 100 & 0.915 & 1.64 & 0.694 & 4.3$^{+1.7}_{-1.2}$ & 11.12$\pm$2.50 & 14.06$\pm$3.17 & 32.42$\pm$7.28 \\      
\texttt{XLSSC} 101 & 0.756 & 1.50 & 0.788 & 4.6$^{+0.8}_{-0.8}$ & 12.27$\pm$0.96 & 16.53$\pm$1.29 & 39.95$\pm$3.13 \\      
\texttt{XLSSC} 102 & 0.969 & 1.69 & 0.574 & 3.2$^{+0.8}_{-0.5}$ & 13.31$\pm$1.41 & 16.07$\pm$1.70 & 33.56$\pm$3.56 \\     
\texttt{XLSSC} 103 & 0.233 & 1.12 & 0.913 & 3.5$^{+1.2}_{-0.8}$ & 0.90$\pm$0.10  & 1.30$\pm$0.14  & 3.20$\pm$0.35 \\   
\texttt{XLSSC} 104 & 0.294 & 1.15 & 1.038 & 4.7$^{+1.5}_{-1.0}$ & 0.86$\pm$0.10  & 1.36$\pm$0.15  & 4.36$\pm$0.42 \\   
\texttt{XLSSC} 105 & 0.429 & 1.24 & 1.024 & 5.2$^{+1.1}_{-0.8}$ & 7.91$\pm$0.57  & 12.39$\pm$0.89 & 34.34$\pm$2.47 \\     
\texttt{XLSSC} 106 & 0.300 & 1.16 & 0.856 & 3.3$^{+0.4}_{-0.3}$ & 3.16$\pm$0.15  & 4.44$\pm$0.21  & 10.48$\pm$0.49 \\    
\texttt{XLSSC} 107 & 0.436 & 1.25 & 0.711 & 2.7$^{+0.4}_{-0.4}$ & 3.82$\pm$0.32  & 4.89$\pm$0.41  & 10.33$\pm$0.88 \\    
\texttt{XLSSC} 108 & 0.254 & 1.13 & 0.705 & 2.2$^{+0.3}_{-0.2}$ & 1.49$\pm$0.10  & 1.90$\pm$0.13  & 3.86$\pm$0.26 \\   
\texttt{XLSSC} 109 & 0.491 & 1.29 & 0.787 & 3.4$^{+1.3}_{-0.8}$ & 4.71$\pm$0.78  & 6.30$\pm$1.04  & 14.34$\pm$2.37 \\     
\texttt{XLSSC} 110 & 0.445 & 1.25 & 0.525 & 1.6$^{+0.1}_{-0.1}$ & 1.43$\pm$0.22  & 1.63$\pm$0.25  & 2.82$\pm$0.43 \\   
\texttt{XLSSC} 111 & 0.299 & 1.16 & 1.017 & 4.5$^{+0.6}_{-0.5}$ & 4.27$\pm$0.21  & 6.65$\pm$0.32  & 18.06$\pm$0.87 \\     
\texttt{XLSSC} 112 & 0.139 & 1.07 & 0.653 & 1.8$^{+0.2}_{-0.2}$ & 0.49$\pm$0.06  & 0.61$\pm$0.08  & 1.15$\pm$0.15 \\   
\texttt{XLSSC} 501 & 0.333 & 1.18 & 0.768 & 2.8$^{+0.6}_{-0.4}$ & 1.84$\pm$0.22  & 2.44$\pm$0.29  & 5.34$\pm$0.64 \\   
\texttt{XLSSC} 502 & 0.141 & 1.07 & 0.532 & 1.2$^{+0.0}_{-0.1}$ & 0.55$\pm$0.04  & 0.63$\pm$0.05  & 1.00$\pm$0.08 \\   
\texttt{XLSSC} 503 & 0.336 & 1.18 & 0.642 & 2.0$^{+0.3}_{-0.2}$ & 2.01$\pm$0.19  & 2.47$\pm$0.24  & 4.79$\pm$0.46 \\  
\end{tabular}                                
\hspace{0.01cm}
\end{center}
\end{table*}    

\begin{table*}
\begin{center}
\contcaption{{\em continued....}}
%\vspace{-0.3cm}
\begin{tabular}{lccccccc}
\hline\hline
\texttt{XLSSC} Num & z & E(z) & $r_{500,MT}$ & $T_{\rm 300kpc}$ & $L^{\rm
  XXL}_{\rm 300kpc}$ & $L^{\rm XXL}_{500,MT}$ & $L^{\rm bol}_{500,MT}$ \\  
 & & & Mpc & (keV) & 10$^{43}$(ergs s$^{-1}$) & 10$^{43}$(ergs s$^{-1}$) & 10$^{43}$(ergs s$^{-1}$) \\
\hline
\texttt{XLSSC} 504 & 0.243 & 1.12 & 1.953 & 13.8$^{+-13.8}_{-5.4}$ & 0.48$\pm$0.17 & 1.35$\pm$0.48 & 6.87$\pm$2.46 \\
\texttt{XLSSC} 505 & 0.055 & 1.02 & 0.661 & 1.7$^{+0.2}_{-0.1}$ & 0.38$\pm$0.02  & 0.47$\pm$0.03  & 0.90$\pm$0.05 \\    
\texttt{XLSSC} 506 & 0.717 & 1.49 & 0.798 & 4.5$^{+2.1}_{-1.5}$ & 6.31$\pm$1.25  & 8.53$\pm$1.69  & 20.69$\pm$4.08 \\       
\texttt{XLSSC} 507 & 0.566 & 1.34 & 0.612 & 2.4$^{+0.6}_{-0.5}$ & 3.68$\pm$0.63  & 4.43$\pm$0.76  & 8.77$\pm$1.50 \\     
\texttt{XLSSC} 508 & 0.539 & 1.32 & 0.742 & 3.3$^{+0.7}_{-0.5}$ & 3.48$\pm$0.33  & 4.55$\pm$0.43  & 10.08$\pm$0.95 \\     
\texttt{XLSSC} 509 & 0.633 & 1.39 & 0.806 & 4.2$^{+1.1}_{-0.8}$ & 6.61$\pm$0.64  & 8.99$\pm$0.86  & 21.49$\pm$2.07 \\      
\texttt{XLSSC} 510 & 0.394 & 1.22 & 0.711 & 2.6$^{+0.4}_{-0.3}$ & 2.31$\pm$0.16  & 2.96$\pm$0.20  & 6.22$\pm$0.43 \\    
\texttt{XLSSC} 511 & 0.130 & 1.06 & 0.545 & 1.3$^{+0.1}_{-0.1}$ & 0.25$\pm$0.03  & 0.29$\pm$0.04  & 0.48$\pm$0.06 \\    
\texttt{XLSSC} 512 & 0.402 & 1.22 & 0.848 & 3.6$^{+0.6}_{-0.4}$ & 2.14$\pm$0.14  & 2.99$\pm$0.19  & 7.11$\pm$0.46 \\    
\texttt{XLSSC} 513 & 0.378 & 1.21 & 0.936 & 4.2$^{+0.8}_{-0.5}$ & 4.40$\pm$0.30  & 6.50$\pm$0.44  & 16.63$\pm$1.12 \\       
\texttt{XLSSC} 514 & 0.169 & 1.08 & 0.582 & 1.5$^{+0.2}_{-0.1}$ & 0.40$\pm$0.06  & 0.47$\pm$0.07  & 0.81$\pm$0.12 \\    
\texttt{XLSSC} 515 & 0.101 & 1.05 & 0.540 & 1.2$^{+0.1}_{-0.1}$ & 0.32$\pm$0.03  & 0.37$\pm$0.04  & 0.59$\pm$0.06 \\    
\texttt{XLSSC} 516 & 0.866 & 1.60 & 0.695 & 4.8$^{+1.0}_{-0.7}$ & 18.38$\pm$1.97 & 23.31$\pm$2.50 & 55.21$\pm$5.91 \\      
\texttt{XLSSC} 517 & 0.699 & 1.45 & 0.698 & 3.4$^{+1.1}_{-0.6}$ & 5.77$\pm$0.88  & 7.33$\pm$1.12  & 16.09$\pm$2.47 \\      
\texttt{XLSSC} 518 & 0.177 & 1.09 & 0.535 & 1.3$^{+0.0}_{-0.0}$ & 0.50$\pm$0.05  & 0.58$\pm$0.05  & 0.93$\pm$0.09 \\    
\texttt{XLSSC} 519 & 0.270 & 1.14 & 0.555 & 1.5$^{+0.2}_{-0.2}$ & 0.81$\pm$0.15  & 0.94$\pm$0.18  & 1.59$\pm$0.30 \\     
\texttt{XLSSC} 520 & 0.175 & 1.08 & 0.805 & 2.7$^{+0.2}_{-0.1}$ & 1.72$\pm$0.05  & 2.34$\pm$0.07  & 5.19$\pm$0.16 \\    
\texttt{XLSSC} 521 & 0.807 & 1.54 & 0.775 & 4.7$^{+1.3}_{-0.8}$ & 12.99$\pm$1.46  & 17.32$\pm$1.95 & 42.03$\pm$4.72 \\      
\texttt{XLSSC} 522 & 0.395 & 1.22 & 0.711 & 2.6$^{+0.4}_{-0.3}$ & 2.11$\pm$0.15  & 2.71$\pm$0.19  & 5.69$\pm$0.40 \\     
\texttt{XLSSC} 523 & 0.343 & 1.18 & 0.779 & 2.9$^{+0.6}_{-0.4}$ & 2.17$\pm$0.17  & 2.90$\pm$0.23  & 6.40$\pm$0.51 \\    
\texttt{XLSSC} 524 & 0.270 & 1.14 & 0.754 & 2.6$^{+0.5}_{-0.4}$ & 0.92$\pm$0.09  & 1.21$\pm$0.12  & 2.59$\pm$0.25 \\    
\texttt{XLSSC} 525 & 0.379 & 1.21 & 0.832 & 3.4$^{+0.3}_{-0.3}$ & 4.83$\pm$0.24  & 6.68$\pm$0.33  & 15.54$\pm$0.76 \\      
\texttt{XLSSC} 526 & 0.273 & 1.14 & 0.794 & 2.8$^{+0.4}_{-0.2}$ & 3.91$\pm$0.20  & 5.27$\pm$0.27  & 11.68$\pm$0.60 \\    
\texttt{XLSSC} 527 & 0.076 & 1.03 & 0.926 & 3.1$^{+2.8}_{-1.0}$ & 0.14$\pm$0.03  & 0.20$\pm$0.05  & 0.50$\pm$0.12 \\    
\texttt{XLSSC} 528 & 0.302 & 1.16 & 0.839 & 3.2$^{+0.8}_{-0.4}$ & 1.49$\pm$0.11  & 2.07$\pm$0.16  & 4.82$\pm$0.36 \\    
\texttt{XLSSC} 529 & 0.547 & 1.33 & 0.769 & 3.5$^{+0.7}_{-0.4}$ & 5.20$\pm$0.44  & 6.91$\pm$0.58  & 15.71$\pm$1.32 \\      
\texttt{XLSSC} 530 & 0.182 & 1.09 & 0.686 & 2.0$^{+0.2}_{-0.2}$ & 0.60$\pm$0.05  & 0.75$\pm$0.06  & 1.49$\pm$0.13 \\    
\texttt{XLSSC} 531 & 0.391 & 1.22 & 0.966 & 4.5$^{+2.2}_{-1.4}$ & 1.81$\pm$0.25  & 2.73$\pm$0.38  & 7.18$\pm$0.99 \\    
\texttt{XLSSC} 532 & 0.392 & 1.22 & 0.772 & 3.0$^{+0.6}_{-0.5}$ & 2.58$\pm$0.23  & 3.43$\pm$0.31  & 7.59$\pm$0.68 \\    
\texttt{XLSSC} 533 & 0.107 & 1.05 & 0.789 & 2.4$^{+0.1}_{-0.1}$ & 1.64$\pm$0.04  & 2.21$\pm$0.05  & 4.80$\pm$0.12 \\    
\texttt{XLSSC} 534 & 0.853 & 1.58 & 0.725 & 4.3$^{+1.7}_{-1.0}$ & 12.49$\pm$1.87 & 16.14$\pm$2.41 & 37.74$\pm$5.66 \\      
\texttt{XLSSC} 535 & 0.172 & 1.08 & 0.756 & 2.4$^{+0.3}_{-0.2}$ & 1.83$\pm$0.10  & 2.41$\pm$0.13  & 5.14$\pm$0.28 \\    
\texttt{XLSSC} 536 & 0.170 & 1.08 & 0.659 & 1.8$^{+0.3}_{-0.2}$ & 0.38$\pm$0.06  & 0.47$\pm$0.08  & 0.91$\pm$0.14 \\     
\texttt{XLSSC} 537 & 0.515 & 1.30 & 0.934 & 4.8$^{+1.2}_{-0.9}$ & 5.47$\pm$0.45  & 8.07$\pm$0.67  & 21.11$\pm$1.76 \\      
\texttt{XLSSC} 538 & 0.332 & 1.18 & 0.804 & 3.1$^{+0.9}_{-0.6}$ & 1.35$\pm$0.14  & 1.83$\pm$0.19  & 4.13$\pm$0.42 \\     
\texttt{XLSSC} 539 & 0.184 & 1.09 & 0.520 & 1.2$^{+0.1}_{-0.2}$ & 0.38$\pm$0.07  & 0.44$\pm$0.08  & 0.69$\pm$0.12 \\    
\texttt{XLSSC} 540 & 0.414 & 1.23 & 0.776 & 3.1$^{+0.4}_{-0.4}$ & 4.13$\pm$0.25  & 5.52$\pm$0.34  & 12.31$\pm$0.75 \\      
\texttt{XLSSC} 541 & 0.188 & 1.09 & 0.805 & 2.7$^{+0.3}_{-0.3}$ & 1.05$\pm$0.06  & 1.42$\pm$0.09  & 3.16$\pm$0.20 \\     
\texttt{XLSSC} 542 & 0.402 & 1.22 & 1.202 & 6.8$^{+0.5}_{-0.3}$ & 28.69$\pm$0.60 & 50.37$\pm$1.05 & 157.70$\pm$4.59 \\       
\texttt{XLSSC} 543 & 0.381 & 1.21 & 0.689 & 2.4$^{+0.5}_{-0.3}$ & 1.06$\pm$0.14  & 1.33$\pm$0.18  & 2.73$\pm$0.36 \\    
\texttt{XLSSC} 544 & 0.095 & 1.04 & 0.788 & 2.4$^{+0.2}_{-0.2}$ & 0.57$\pm$0.02  & 0.77$\pm$0.03  & 1.68$\pm$0.07 \\    
\texttt{XLSSC} 545 & 0.353 & 1.19 & 0.668 & 2.2$^{+1.6}_{-0.6}$ & 1.13$\pm$0.33  & 1.41$\pm$0.41  & 2.82$\pm$0.81 \\    
\texttt{XLSSC} 546 & 0.792 & 1.53 & 0.668 & 3.5$^{+0.7}_{-0.6}$ & 10.49$\pm$1.09 & 13.08$\pm$1.36 & 28.47$\pm$2.97 \\  
\texttt{XLSSC} 547 & 0.371 & 1.20 & 0.920 & 4.0$^{+1.1}_{-0.8}$ & 2.80$\pm$0.27  & 4.09$\pm$0.40  & 10.31$\pm$1.00 \\          
\texttt{XLSSC} 548 & 0.321 & 1.18 & 0.428 & 1.0$^{+0.1}_{-0.1}$ & 0.47$\pm$0.13  & 0.51$\pm$0.13  & 0.74$\pm$0.19 \\    
\texttt{XLSSC} 549 & 0.808 & 1.54 & 0.709 & 4.0$^{+2.4}_{-0.9}$ & 8.87$\pm$1.50  & 11.34$\pm$1.92 & 25.92$\pm$4.40 \\   
\hline                                
\end{tabular}        
\hspace{0.01cm}                       
\end{center}                          
\end{table*}

\subsection{Sample $LT$ Relation}
\label{sec:ltunbias}

\begin{figure*}
\centering
\includegraphics[width=9.1cm]{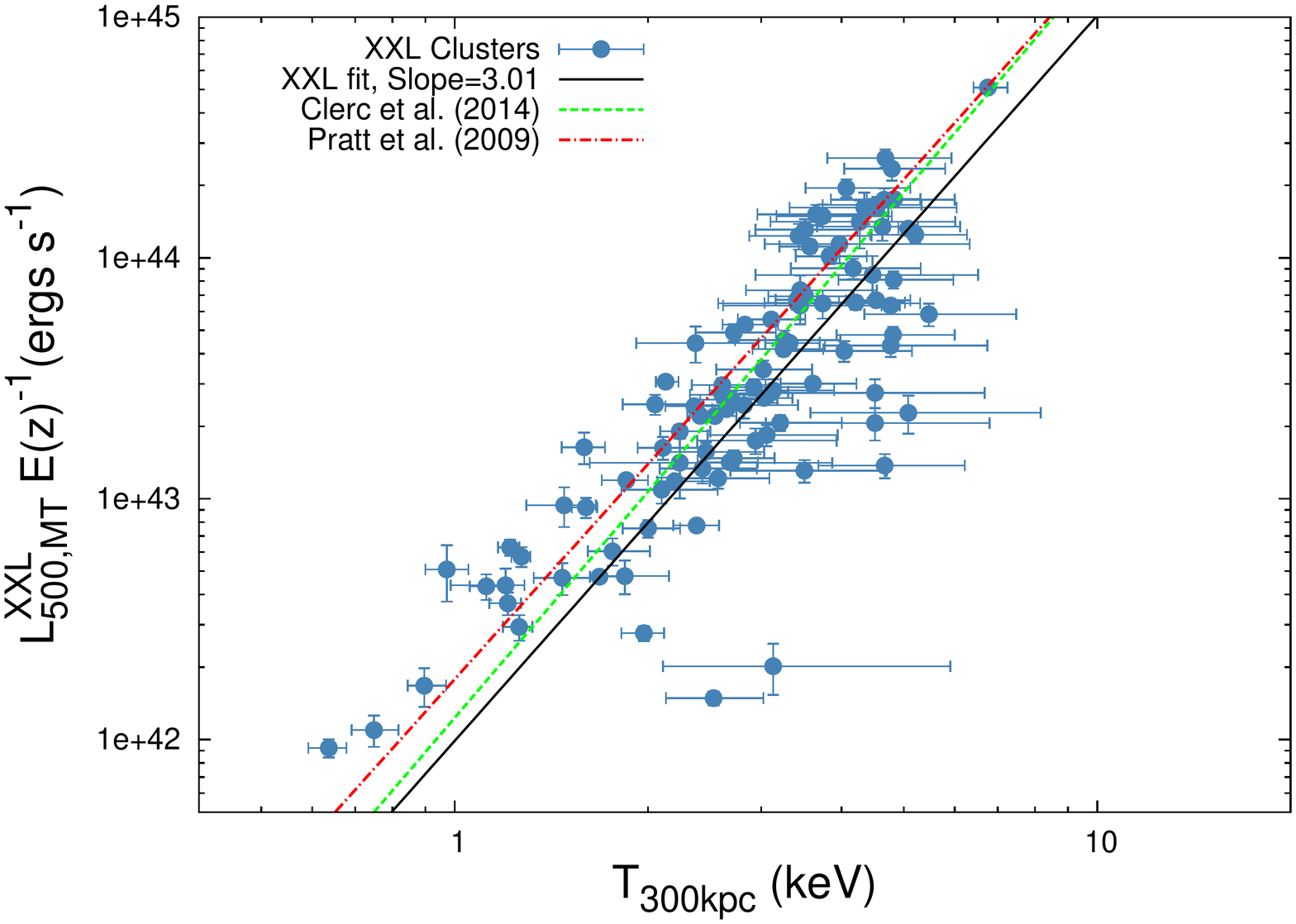}
\includegraphics[width=9.1cm]{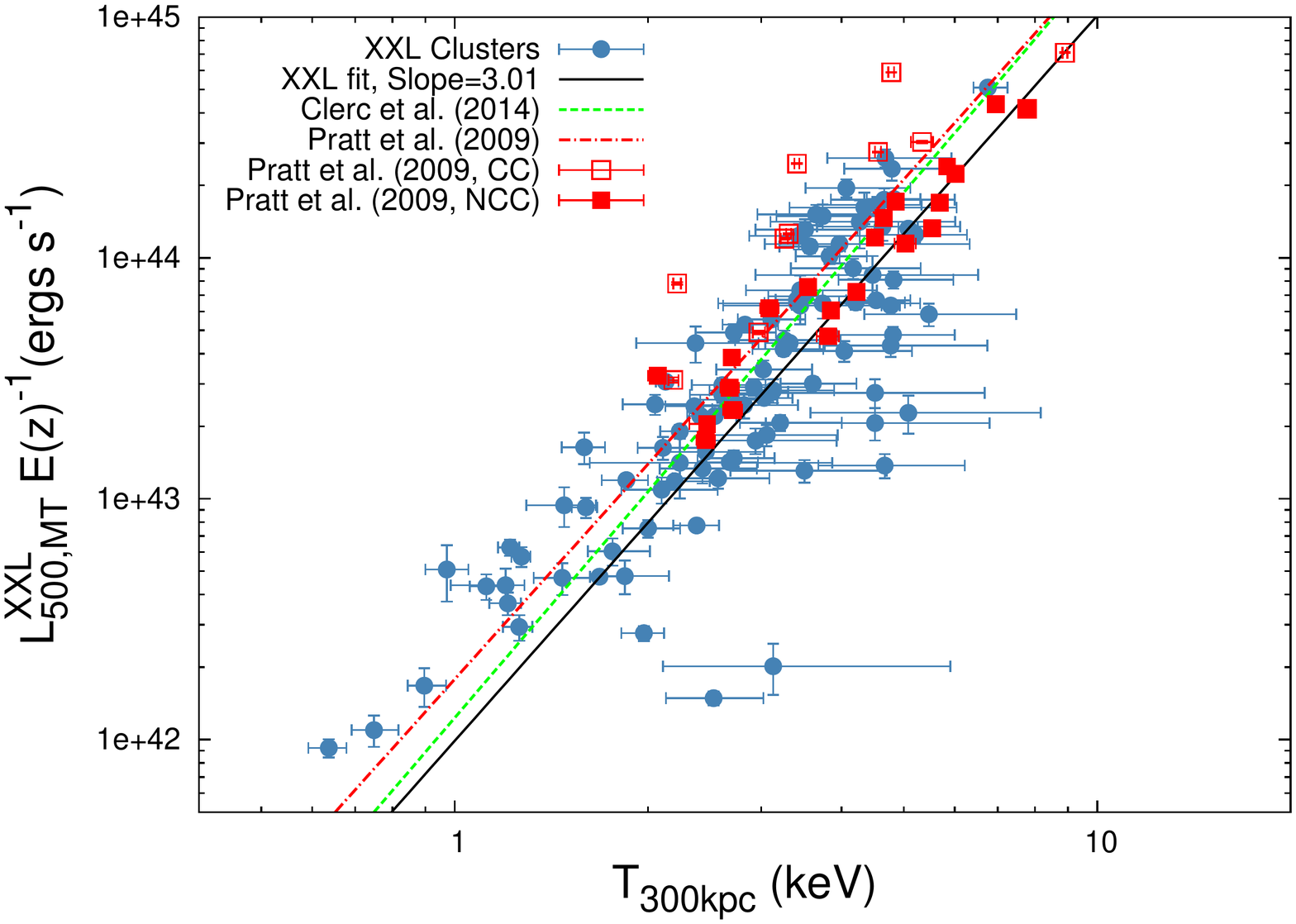}
\vspace{-1cm}
\caption[]{\small{(Left) $LT$ relation for the XXL sample.  The black line
    represents an unbiased fit to the data (see
    Sect.~\ref{sec:ltunbias}).  The $LT$ relations of
    \cite{2009A&A...498..361P} and \cite{2014MNRAS.444.2723C}, given
    by the red dot-dashed line and the green dashed line,
    respectively, are overplotted; (Right) Same as the left plot, but
    with the REXCESS clusters studied in \cite{2009A&A...498..361P}
    overplotted, split between cool-core (open red squares) and
    non-cool-core (filled red squares) clusters.  In both
    plots, the errors on the temperatures for the XXL clusters have
    been transformed via the method described in
    Sect.~\ref{sec:results}.}\label{fig:ltunbias}}
\end{figure*}

Figure~\ref{fig:ltunbias} (left panel) shows the $\lxxl-T$ relation for the 
XXL-100-GC.  For simplicity, the $\lxxl-T$ notation
explicity refers to the $L^{\rm XXL}_{500,MT}-T_{\rm 300kpc}$ relation.  A fit to
the data using a power law of the form 
\begin{align}
\hspace{2.5cm}\left(\frac{\lxxl}{L_{0}}\right) &= E(z)^{\gamma_{LT}}A_{\rm LT}\left(\frac{T}{T_{0}}\right)^{B_{\rm LT}}
\label{equ:ltunbias}
\end{align}
was performed, where $A_{LT}$, $B_{LT}$, and $\gamma_{LT}$ represent the
normalisation, slope, and power of the evolution correction
respectively.  We note that the two clusters at low luminosity offset
from the $\lxxl-T$ are the clusters \texttt{XLSSC} 011 and 527.  These clusters
are low S/N clusters at low redshift, 0.054 and 0.076, such that the
300 kpc region is large on the detector.  The temperature measurement
is likely affected by these factors.  However, we still include them
in the fit to the $\lxxl-T$ relation.  The power law was fit to
the data using the BCES orthogonal regression in base ten log space
\citep{1996ApJ...470..706A} assuming self-similar evolution.  The
fit is given by the black solid line in Figure~\ref{fig:ltunbias},
assuming $L_{0}$=3$\times$10$^{43}$ erg s$^{-1}$,
$T_{0}$=3keV, and $\gamma_{LT}$=1 (self-similar).  We find a
normalisation of A$_{\rm LT}$=0.90$\pm$0.06, and slope 
B$_{\rm LT}$=3.03$\pm$0.27.  For comparison, we fit the $LT$ relation
to the 31 REXCESS clusters studied in \citet[][hereafter
  P09]{2009A&A...498..361P}, and 52 clusters selected within the
11deg$^{2}$ {\em XMM}-LSS survey 
\citep[][hereafter C14]{2014MNRAS.444.2723C}.  To fit for these $LT$
relations we use the data from Table B1 (Cols. $T_{1}$ and
$L[0.5-2]_{1}$) in P09, and the data given in Table 1 (Cols. $T_{X}$
and $L^{[0.5-2]}_{500}$) in C14.  The P09 and C14 relations are hereafter
denoted as $L_{P09}-T$ and $L_{C14}-T$, respectively, and the fit
parameters are given in Table~\ref{tab.lt}.  The fits to the P09 and C14
clusters are given by the red dot-dashed line and green dashed line,
respectively, assuming the same $L_{0}$, $T_{0}$, and $\gamma_{LT}$ as
for the XXL-100-GC fit.  We find no significant difference between the
XXL-100-GC $\lxxl-T$ relation and the $L_{C14}-T$ relation.  This is
unsurprising as the C14 clusters are selected from the {\em XMM}-LSS
area, which has many clusters in common with the XXL-100-GC.  

We find a 2.8$\sigma$ difference in the normalisation as compared to
the REXCESS clusters, with the XXL normalisation being lower.  This is
due to the presence of strong cool core clusters in the REXCESS
sample.  These clusters are apparent as high-luminosity outliers in
Fig.~\ref{fig:ltunbias} (right), which shows the REXCESS clusters plotted
on the XXL $LT$ relation.  Qualitatively, it is clear that in the
absence of these strong cool core clusters, the remaining REXCESS
clusters are consistent with the $\lxxl-T$ relation. As discussed in
section~\ref{sec:coolcore}, the absence of strong cool core clusters from
the XXL-100-GC sample is due to a combination of survey geometry and
cool core evolution.           
 
\subsection{Selection function}
\label{sec:sfunc}

Full details of the construction of the selection function are given
in Paper II, but the key points are summarised here.  The
selection function takes into account three aspects of the XXL-100-GC
selection, (i) the pipeline detection, (ii) the flux cut of the
XXL-100-GC, and (iii) the survey sensitivity. The pipeline
detection, i.e. the C1+2 classification in pointing $p$ (denoted
$P_{C1+2,p_{i}}(I|CR_{\infty},r_{c},RA,Dec)$), was studied in depth in
\cite{2006MNRAS.372..578P} and updated following the methodology of
\cite{2012MNRAS.423.3561C}.  We refer the reader to these works for
full details.  The modelling of the flux cut assumes that the errors on
the GCA count rate can be modelled as a Gaussian distribution.  The
probability of a true aperture count rate (CR$_{60}$) being greater
that the count-rate cut (CR$_{\rm cut}$) for a given pointing
($p_{i}$) is 
\begin{align}
P_{P_{i}}(I|CR_{\infty},r_{c},RA,& Dec) =
\frac{P_{C1+2,p_{i}}(I|CR_{\infty},r_{c},RA,Dec)}{2} \nonumber \\ \times
 & \left(1 + {\rm erf}\left[\frac{\epsilon_{60}(r_{c})CR_{\infty} - CR_{cut}}{\sigma_{m}(\epsilon_{60}(r_{c})CR_{\infty},RA,Dec)\sqrt{2}}\right]\right),
\label{equ:sfunc}
\end{align} 
where $\epsilon_{60}(r_{c})=\left(1 -
  \left[1 + \left(60^{\prime\prime}/r_{c}\right)^{2}\right]^{1.5-3\beta}\right)$ and CR$_{\infty}$ is
the pipeline count rate extrapolated to infinity.  The final
ingredient is the modelling of pointing overlaps, where we assume the
detection is independent over the different pointings.  For a given
position, pointings are sorted by increasing off-axis distance,
reproducing the overlap cross-matching procedure of the survey.  The
combined selection function is then given by
\begin{align}
\hspace{1.4cm}
P(I) & = \sum_{i=1}^{N}\left[P_{p_{i}}(I)\prod_{j<i}\left(1 - P_{C1+2,p_{j}}(I)\right)\right].
\end{align}

\subsection{Likelihood}
\label{sec-1}

The observational data used in this study is the distribution of the
XXL-100-GC clusters in $L$, $T$, and $z$. Our physical model assumes
that the cluster population is described by a power-law correlation
between $L$ and $T$ with log-normal scatter in $L$ and  evolution in
$L$ by $E(z)^{\gLT}$. The XXL-100-GC data represent a subset of this
population selected according to our selection function and with noisy
measurements of $L$ and $T$, denoted $\Lh$ and $\Th$. We neglect any
measurement errors on $z$.   

The number density of clusters in the survey volume is taken into
account by using a mass function, assumed to be described by a Tinker
mass function \citep{2008ApJ...688..709T}. We then transform
this to a temperature function, $dn/dT$, using the $M_{WL}-T$ relation
given in Sect.~\ref{sec:analysis}.  In the present analysis we
neglect the intrinsic scatter in the $M_{WL}-T$ relation, and the
measurement errors on the parameters of the scaling relation, and
assume that its evolution is self-similar (i.e. $M \propto E(z) T^{B_{MT}}$).

The total number of clusters $N$ in the volume is then the integral
over the temperature and redshift range considered, multiplied by the
solid angle of the survey $\Omega$:
\begin{align}
\hspace{2cm}\left< N \right> & = \Omega \int dT \int dz\, \frac{dn}{dT} \frac{dVd\Omega}{dz}.
\end{align}

The number of clusters predicted by our model to be \emph{observed} in
the subsample defined by our selection function is the integral of the
mass function over the volume of the survey, weighted by the
probability that a cluster of a given mass would be included in the
subsample given the $LT$ relation and the intrinsic and statistical
scatter on the luminosity,
\begin{align}
\label{eq.Ndet}
\hspace{1.5cm}\left< N_{det} \right>  & = \int dT \int dz \,
\frac{dn}{dT} \frac{dVd\Omega}{dz}\Omega \nonumber \\
  &  \times \int dL \, P(L|T,z,\theta) \times P(I|L,T,z),
\end{align}
where $\theta$ stands for our full set of model parameters
(\ALT,\BLT,\gLT,\intLT) describing the $LT$ relation. In this
expression, the first probability, $P(L|T,z,\theta),$ is the
probability that cluster of temperature $T$ has some intrinsically
scattered luminosity.  The second probability, $P(I|L,T,z),$ is the
selection function, i.e. the probability that a cluster with a
luminosity $L$ and temperature $T$ at a redshift $z$ would be included
in the subsample. This is summarised in Sect.~\ref{sec:sfunc}.  We
note the change in notation as we are writing the likelihood in terms
of the cluster properties $L$ and $T$.

The (un-normalised) likelihood of a cluster $i$ in our sample having
the observed properties ($\hat{L}_{i}$,$\hat{T}_{i}$) is given by
\begin{align}
\label{eq.pltm}
\hspace{0.8cm}P_i(\hat{L}_{i},\hat{T}_{i},I|z_i,\theta) & = \int dT\, \int dL\, \frac{\frac{dn}{dT}\frac{dVd\Omega}{dz}\Omega}{\left< N \right>} \nonumber \\
 & \times P(L|T,z,\theta) P(\Lh|L) P(\Th|T) P(I|L,T,z).
\end{align}

The inclusion of $\left< N \right>$ normalises the temperature
function to a probability distribution for an arbitrary cluster to
have a temperature $T$ at redshift $z$. The probability of $P(L|T,z)$
is as defined above, and the remaining terms are the probability of
each of the observables using the measured uncertainty for that
observable.  The joint probability of the full set of observed cluster 
properties is the product of $P_i(\Lh,\Th)$ over all $N_{det}$
observed clusters in the sample. 

Equation \ref{eq.pltm} is an improper probability because $P(I|L,T,z)$
does not integrate to unity, and it does not penalise the model for
predicting the existence of clusters in parts of the $L,T,z$ space
where they are within the selection function but are not observed. In
other words, the model is not penalised for excess probability density
in regions of the space where the lack of detections disfavours the
existence of clusters. This is resolved by normalising
Eq. \ref{eq.pltm} by the integral over the observed $\Lh,\Th$ space to
give the final likelihood
\begin{align}
\label{eq.lik}
\hspace{1.6cm}{\cal {L}}(\Lh,\Th,I|z,\theta) & = \prod_i^{N_{det}} \frac{P_i(\Lh,\Th,I|z_i,\theta)}{\int d\Th\, \int d\Lh\, P_i(\Lh,\Th,I|z_i,\theta)}.
\end{align}

This normalisation penalises the model for excess probability density
in regions of the parameter space where the data disfavour the
existence of clusters.  For example, we consider a set of model
parameters that give a good fit to the properties of the observed
clusters, but also give a high probability of cool, highly luminous
clusters, when none are observed. This would lead to a larger
denominator in equation~\ref{eq.pltm}, and hence a lower overall
likelihood compared with an alternate set of parameters that describes
the properties of the observed clusters equally well, but does not
predict the unobserved clusters.  This likelihood is referred to as
the XXL likelihood.

We also consider an alternative construction of the likelihood set
out by M10. In this approach, the final likelihood for the sample of
clusters and their observed properties is the product of a Poisson
likelihood of $N$ total clusters (detected plus undetected)  given the
model prediction $\left< N \right>$, a binomial coefficient accounting
for the number of ways of drawing $N_{det}$ detected clusters from the
total $N$, the joint probability of the set of observed cluster
properties (the product of Eq. \ref{eq.pltm} over the $N_{det}$
clusters), and the probability of not detecting the remaining
$N-N_{det}$ clusters. Neglecting terms not dependent on the model
parameters, the likelihood simplifies to
\begin{align}
\label{eq.likm}
\hspace{1.9cm}{\cal {L}}(\Lh,\Th,I|z,\theta) \propto e^{-\left< N_{det} \right>}\prod_{i=1}^{N_{det}}\left< \tilde{n}_{det,i} \right>,
\end{align}
where
$\left< \tilde{n}_{det,i} \right>=P(\Lh,\Th)\left< N \right>$
for the ith cluster.

The principal difference between Equations \ref{eq.lik} and
\ref{eq.likm} is that the latter has a stronger requirement that the
model must accurately predict the number of observed clusters in
addition to their distribution in ($L,T,z$). This has the advantage
that it uses additional information to constrain the model, and is a
requirement when the analysis is being used to constrain cosmology in
addition to the form of the scaling relations, as in M10.

However, when the aim of the analysis is to infer the form of the
scaling relations under an assumed cosmology, Eq. \ref{eq.lik} has certain
advantages. This implementation is insensitive to systematics
affecting the numbers of clusters (such as the normalisation of the
mass function and hence $\sigma_8$) or the normalisation of the $MT$
relation, and so gives more robust measurements of the scaling
relation parameters.  Furthermore, since the number of observed
clusters is not used to constrain the model parameters in this
approach, the number of detected clusters can be used as a posterior
predictive check.

\subsection{Inference of model parameters}
\label{sec-2}

The likelihood of Eq. \ref{eq.pltm} was combined with priors on each of the
model parameters to compute the posterior probability. The priors used
were uniform in the range ($-\infty,\infty$) for \ALT,\BLT, and \gLT\  and
uniform in the range ($0.01,2.0$) for $\intLT$ (expressed in natural
log, so representing fractional scatter).

The posterior distribution was analysed using the Bayesian inference
package {\em Laplace's Demon}\footnote{\url{http://www.bayesian-inference.com/software}}
within the {\em R} statistical computing environment
\citep{rstat}. The posterior distribution was first explored using a Laplace
approximation for computational efficiency before refining the fit
using an MCMC algorithm. We used the ``Adaptive
Metropolis-within-Gibbs'' algorithm in \emph{Laplace's Demon} for this
purpose, and used four parallel chains  of $50,000$ iterations each,
initialised to randomised starting values (near the mode of the
posterior identified by the Laplace approximation). The stationary
parts of the chains were compared using the Gelman and Rubin (1992)
convergence diagnostic, and the largest value of the $95\%$ upper
bound on the potential scale reduction factor was 1.01, giving a
strong indication that the chains had converged. The stationary parts
of the chains were then concatenated giving an effective sample size
of at least 500 for each parameter.

\subsection{The \lxxl-T Relation}
\label{sec:ltbias}

\begin{figure}[t]
\centering
\includegraphics[angle=0,width=8.2cm]{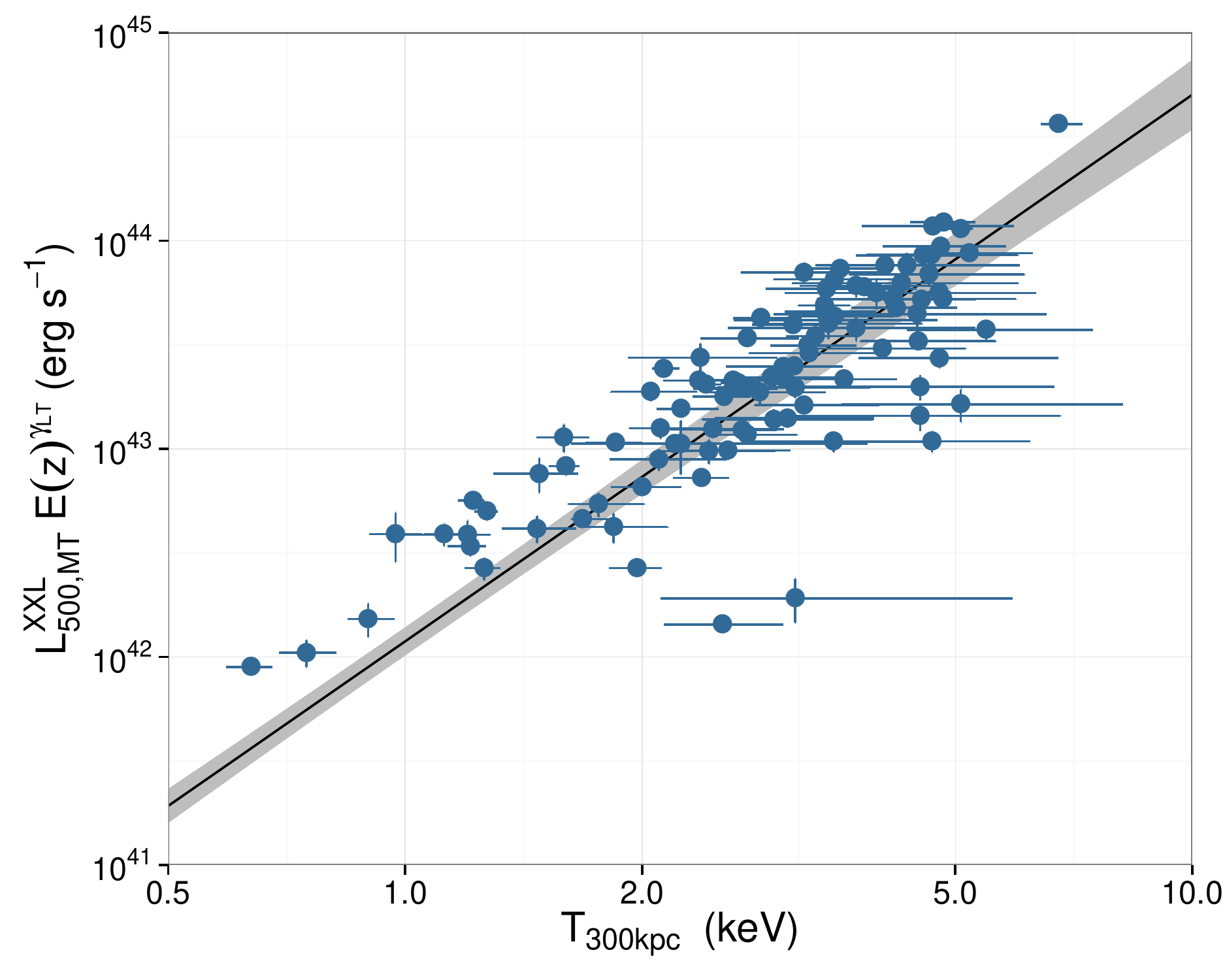}
\caption{\small{\lxxl-$T$ relation with the best-fitting
  model.  The light blue circles show the XXL-100-GC clusters;  the
  best-fitting model is shown as the solid black line the 1$\sigma$
  uncertainty represented by the grey shaded region.}\label{fig:lt}}
\end{figure}

The \lxxl-$T$ relation is plotted in Figure~\ref{fig:lt} along with the
best-fitting model (black line). The best-fitting parameter values and
their uncertainties are then summarised by the mean and standard
deviation of the posterior chains for each parameter. The values are
given in Table~\ref{tab.lt}, and illustrated with the scatterplot
matrix in Fig.~\ref{fig:mat}.  We focus our attention on the XXL
likelihood method, noting that the results do not change significantly
between the XXL and M10 methods.  Using the XXL likelihood method, we
find a normalisation and slope of $A_{LT}$=0.71$\pm$0.11 and
$B_{LT}$=2.63$\pm$0.15, respectively.  We find a  shallower slope than
that found when using the BCES regression fit (which did not account
for biases, see Sect~\ref{sec:ltunbias}), although the difference in
slope is only significant at the 1$\sigma$ level.  The \lxxl-$T$
relation is consistent with recent results which also accounts for
selection biases.  \cite{2015A&A...573A..75B} studied the $LT$
relation of 26 groups (spanning the 0.6 $<$ $T$ $<$ 3.6 keV range) and
found a slope of $B_{LT,B14}\approx2.7$, converting their $LT$
relation from bolometric to the soft band (see
Sect.~\ref{sec:ltbol}).  The comparison is complicated, however,
owing to the differing implementations of the bias-correction and
fitting methods used in \cite{2015A&A...573A..75B} and in our
work. Interestingly, they find that the slope of the sample $LT$
relation is shallower than the bias-corrected fit, opposite to what we
find in this work.  However, this could be driven by the large number
of strong cool core (SCC) systems in their sample ($\approx50\%$);
 the SCC $LT$ relation shows a change in slope from
2.56$\pm$0.22 to 3.60$\pm$0.22 when correcting for biases.               

\begin{figure}[t]
\begin{center}
\includegraphics[width=8.2cm]{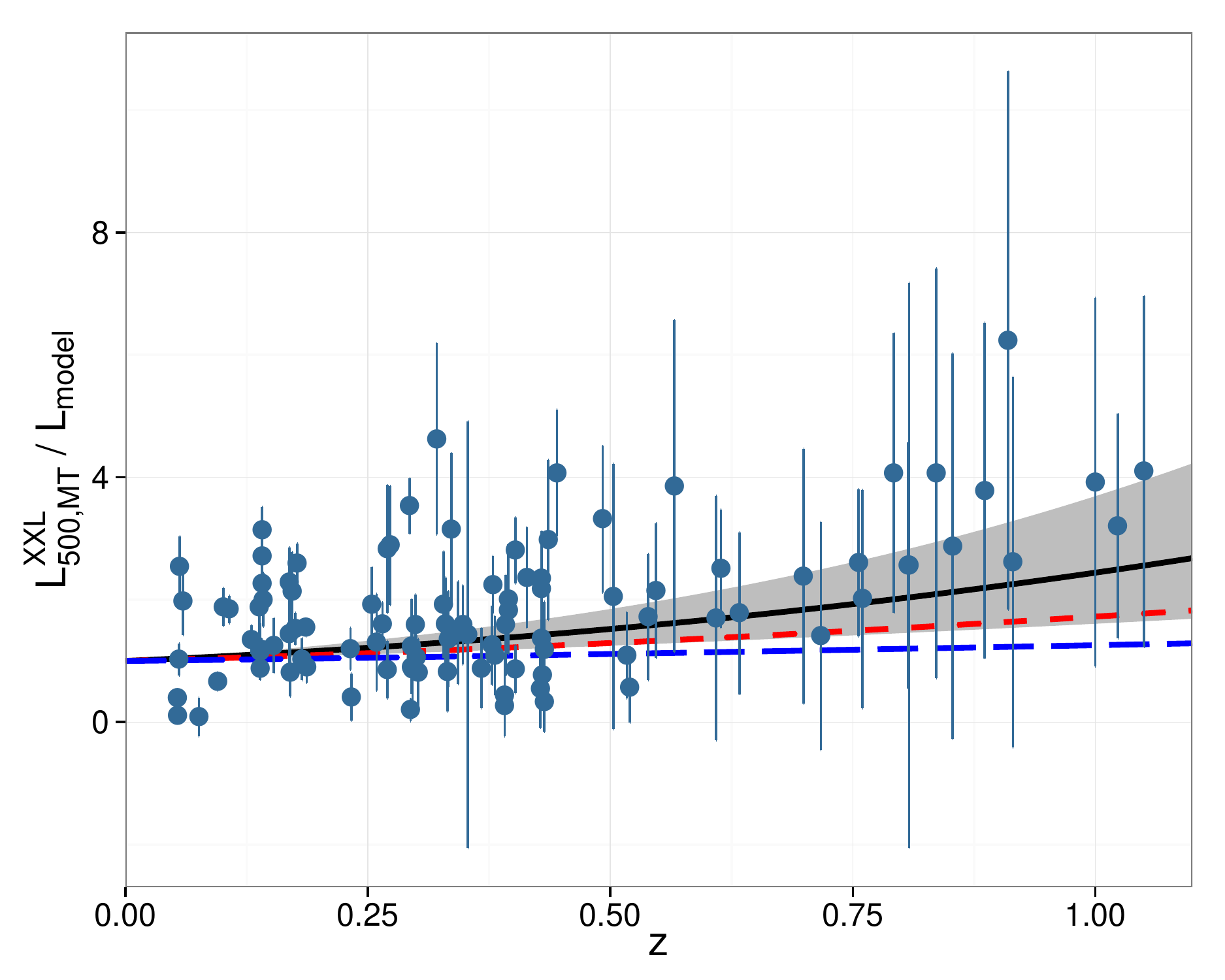}
\end{center}
\caption[]{\small{Evolution of the \lxxl-$T$ relation for the XXL-100-GC.
    The XXL-100-GC are represented by the light blue circles and the best-fitting model is given by the black solid line;  the grey shaded
    region highlights the 1$\sigma$ uncertainty.  The ``strong'' and
  ``weak'' self-similar expectations are given by the red dashed and
    blue dashed lines, respectively.}\label{fig:ltevol}}
\end{figure}

\begin{figure*}[t]
\centering
\includegraphics[angle=0,width=17cm]{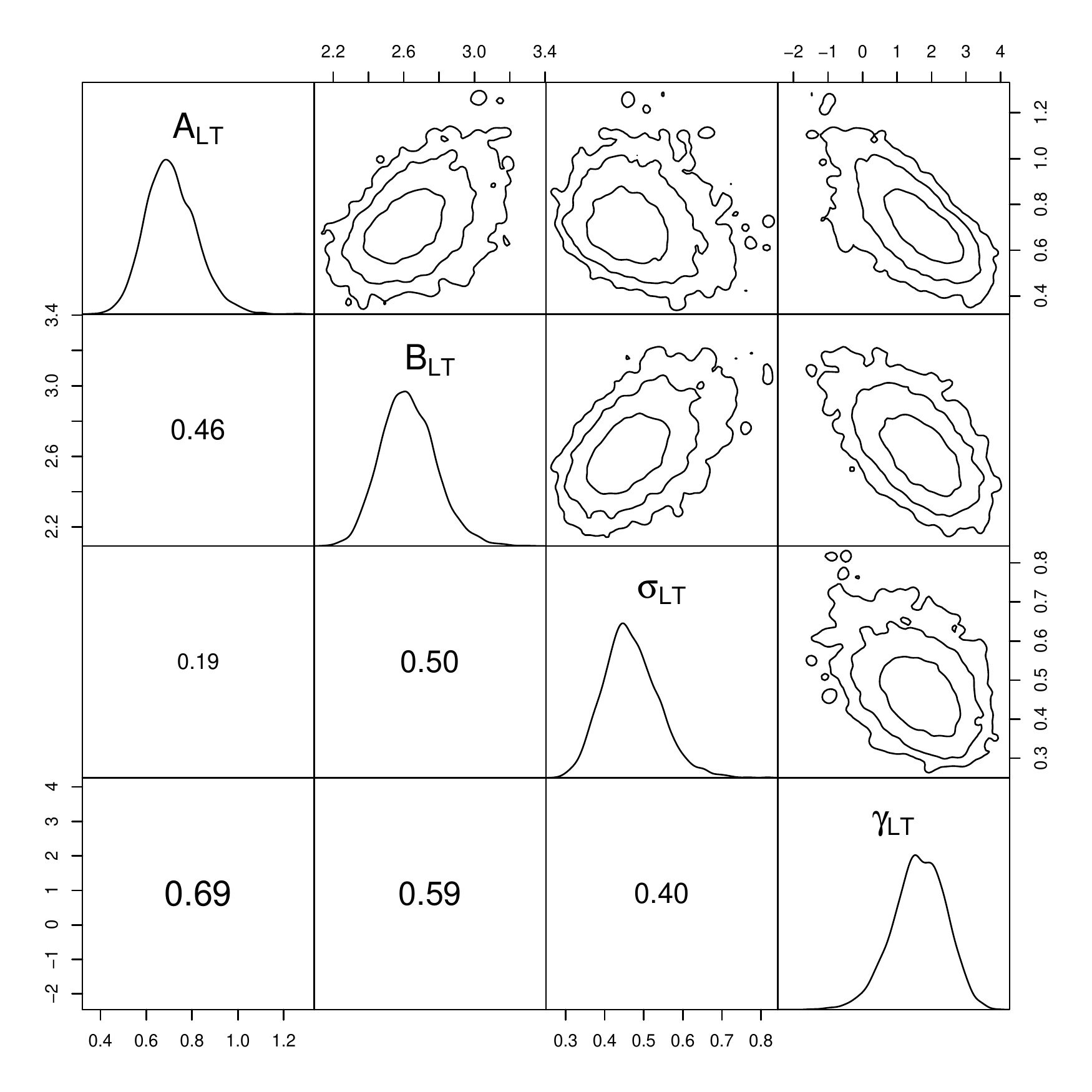}
\caption{\small{Scatterplot matrix for the fit of the
    \lxxl-$T$ relation of the XXL-100-GC sample. The posterior densities are
    shown along the diagonal; the $1\sigma$, $2\sigma$, and $3\sigma$
    confidence contours for the pairs of parameters are shown in the
    upper right panels.  The lower left panels show the Pearson's
    correlation coefficient for the corresponding pair of parameters
    (text size is proportional to the correlation
    strength)}.\label{fig:mat}}
\end{figure*}

Next we focus on the evolution of the \lxxl-$T$  relation, \gLT, where 
the evolution is expressed as $E(z)^{\gLT}$.   For the first time we
are able to measure the evolution of the $LT$ relation using clusters
drawn from a single homogeneous survey, fully accounting for selection
biases.  As introduced previously, we expect \gLT=1 for strong
self-similar evolution, and \gLT=0.42 for weak self-similar
evolution.  We find \gLT=1.64$\pm0.74$, fully consistent with both the
the strong and weak self-similar evolution models.
Figure~\ref{fig:ltevol}  plots the evolution of the $LT$ relation as
inferred from our best-fitting model.  The best-fit evolution is given
by the black solid line along with the 1$\sigma$ uncertainty, the
strong and weak self-similar expectations are given by the red
and blue dashed lines, respectively.  Our best-fit model appears to
lie below all the high-redshift clusters, due to the fit being driven
by the larger number of lower redshift clusters.  However,
the evolution favoured by our best-fit model is in tension with other
recent results, and is discussed further in Sect.~\ref{sec:evol}.       

\begin{table*}
\begin{center}
\begin{tabular}{lcccccc}
  \hline\hline
   Relation & Fit & \ALT\ & \BLT\ & \gLT & \intLT & N$_{\rm det}$\\
   (1) & (2) & (3) & (4) & (5) & (6) & (7) \\
  \hline
  \lxxl-$T$ & BCES & $0.90\pm0.06$ & $3.03\pm0.28$ & 1.0 (fixed) &
  0.53$\pm$0.07 & -- \\
  $L_{P09}$-$T$ & BCES & $1.54\pm0.22$ & $2.97\pm0.34$ & 1.0 (fixed) &
  0.62$\pm$0.07 & -- \\
  $L_{C14}$-$T$ & BCES & $1.26\pm0.25$ & $3.12\pm0.43$ & 1.0 (fixed) &
  0.84$\pm$0.15 & -- \\
  $\mathbf{L_{XXL}}${\bf -}$\mathbf{T}$ & {\bf XXL} &
  $\mathbf{0.71\pm0.11}$ & $\mathbf{2.63\pm0.15}$ &
  $\mathbf{1.64\pm0.77}$ & $\mathbf{0.47\pm0.07}$ & {\bf 117} \\
  \lbol-$T$ & XXL & $1.21\pm0.19$ & $3.08\pm0.15$ & 1.64$\pm$0.77 &
  $0.47\pm0.07$ & -- \\
  \lxxl-$T$ & M10 & $0.71\pm0.10$ & $2.70\pm0.14$ & 1.27$\pm$0.49 &
  $0.50\pm0.06$ & 104 \\
  \hline
\end{tabular}
\caption{\label{tab.lt} Best-fitting parameters for the $LT$ relations
  modelled in this work taking the form
  $L/L_{0}=E(z)^{\gLT}\ALT(T/T_{0})^{\BLT}$, where
  $L_{0}$=3$\times$10$^{43}$ erg s$^{-1}$ and $T_{0}$=3 keV. The fit
  highlighted in bold represents our main result obtained with the bias
  correction method described in Sect~\ref{sec-1}.  The bolometric relations are transformed from the soft-band relation
  via the method described in Sect.~\ref{sec:ltbol}.  (1) $LT$
  relation; (2) fit method; (3) normalisation; (4) slope; (5)
  evolution term ($E(z)^{\gLT}$); (6) intrinsic scatter; and (7) number of
  clusters predicted by the model.}
\end{center}
\end{table*}

\subsection{Posterior predictive checks}
\label{sec-3}

While the modelling process described above determines the
best-fitting values of the model parameters for the chosen model, it
does not guarantee that the model is a good description of the
data. For this we use three types of posterior predictive check to
assess how well the final model describes the data.

\begin{figure*}[t]
\centering
\includegraphics[angle=0,width=6.0cm]{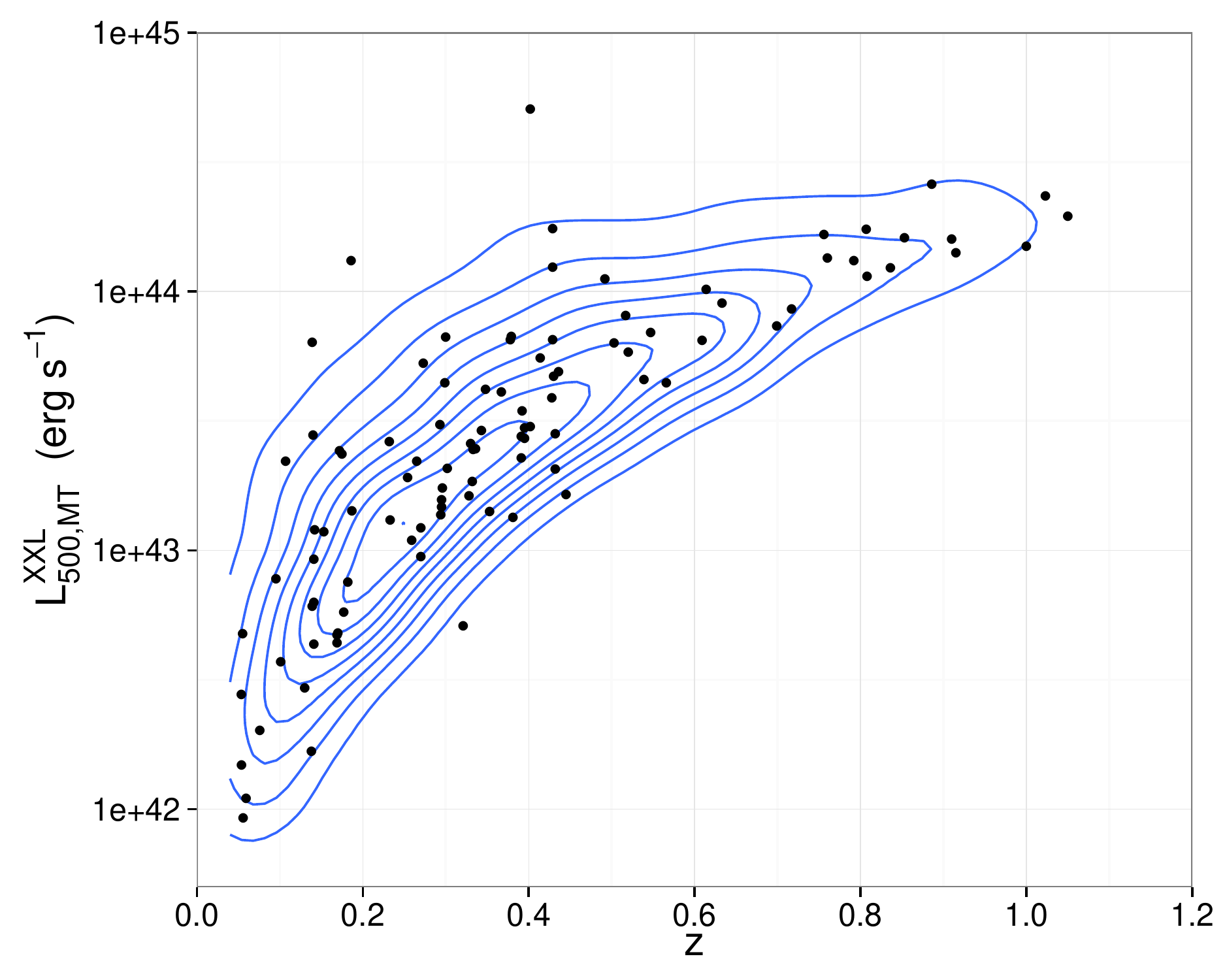}
\includegraphics[angle=0,width=6.0cm]{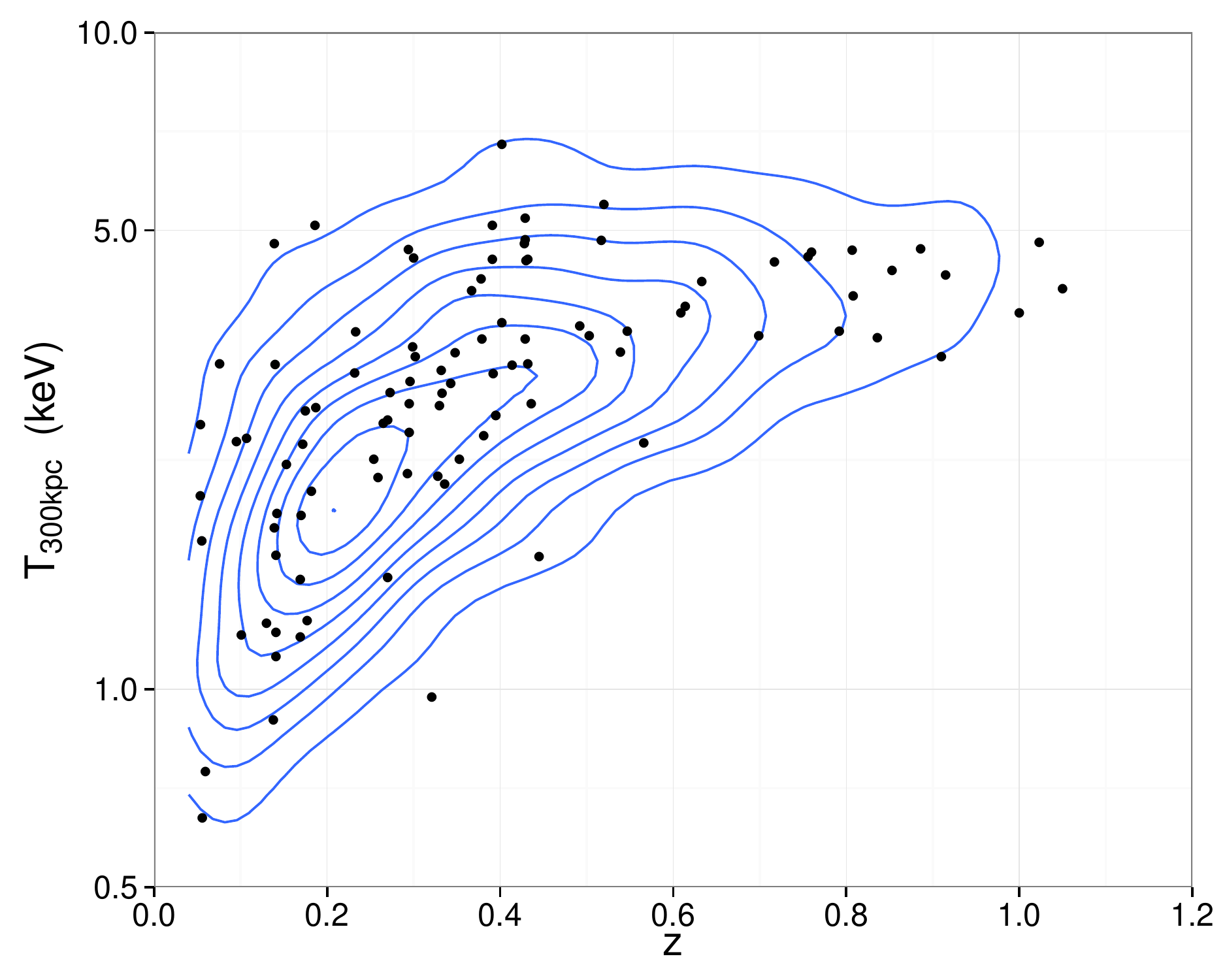}
\includegraphics[angle=0,width=6.0cm]{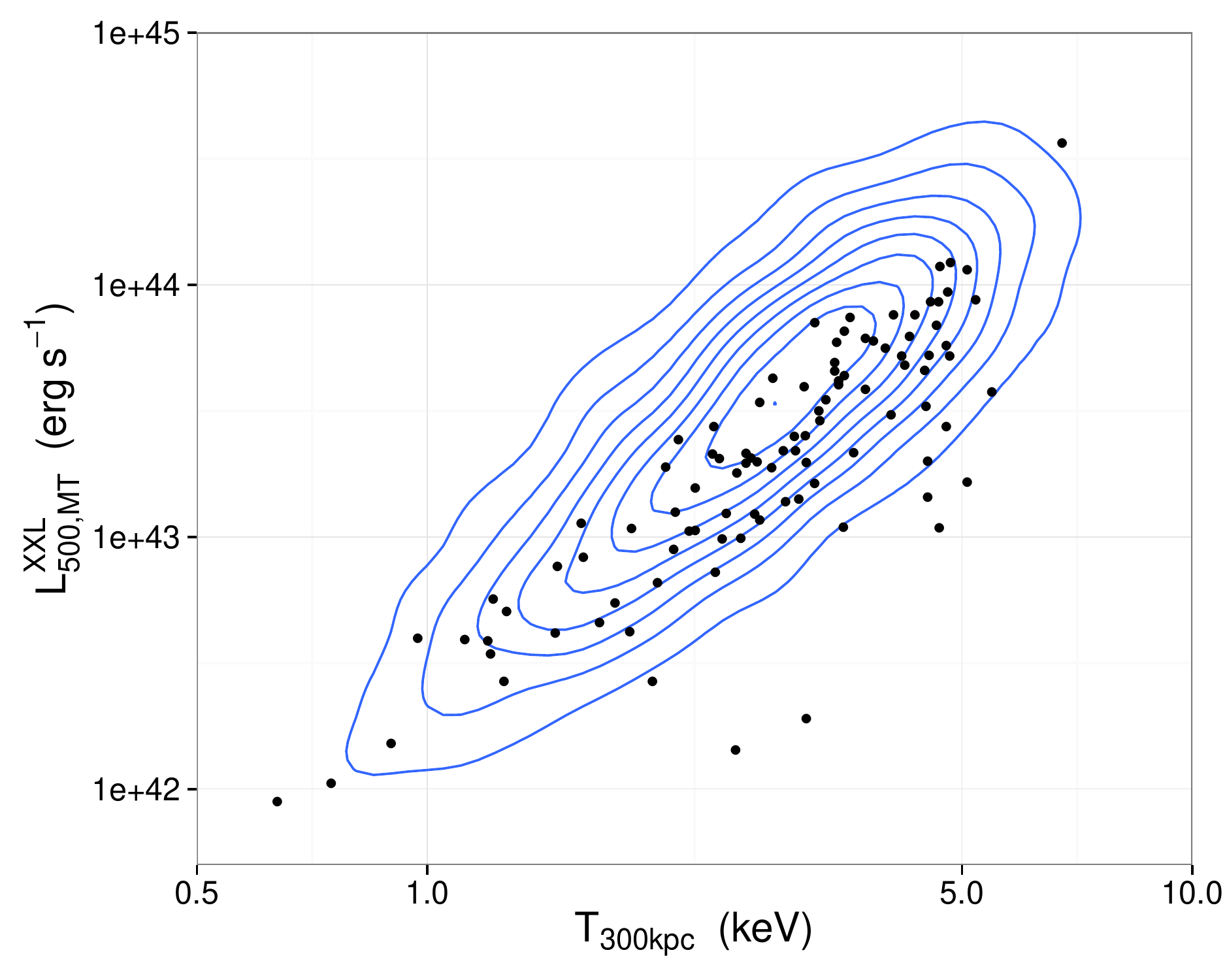}
\caption{\small{Contours of the simulated clusters generated
      from the best-fitting model are plotted in the $L,z$ (left
      panel); $T,z$ (middle panel); and $L,T$ (right panel) planes,
      along with points indicating the observed
      clusters.}\label{fig:cont}}
\end{figure*}

\begin{figure*}[t]
\centering
\includegraphics[angle=0,width=6.0cm]{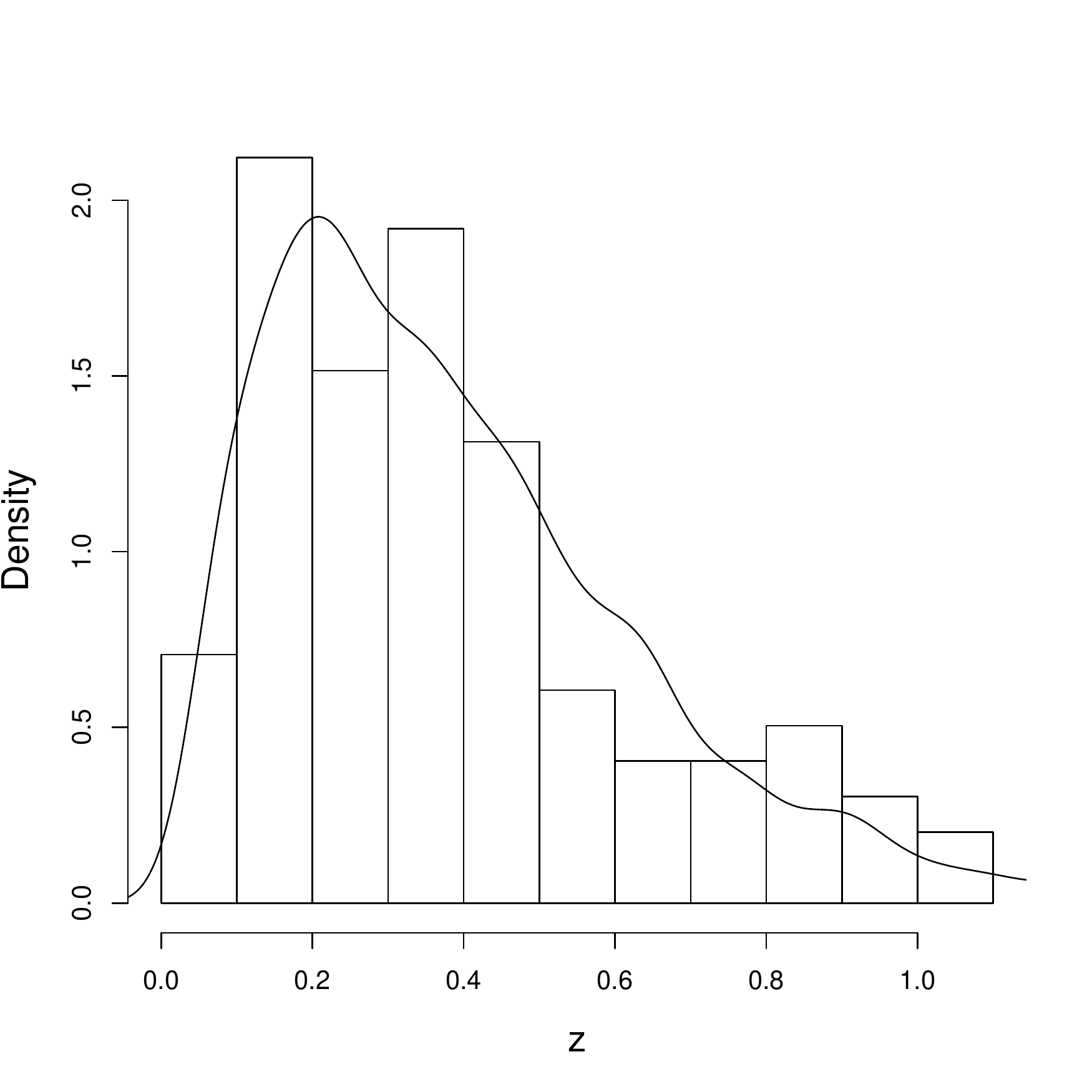}
\includegraphics[angle=0,width=6.0cm]{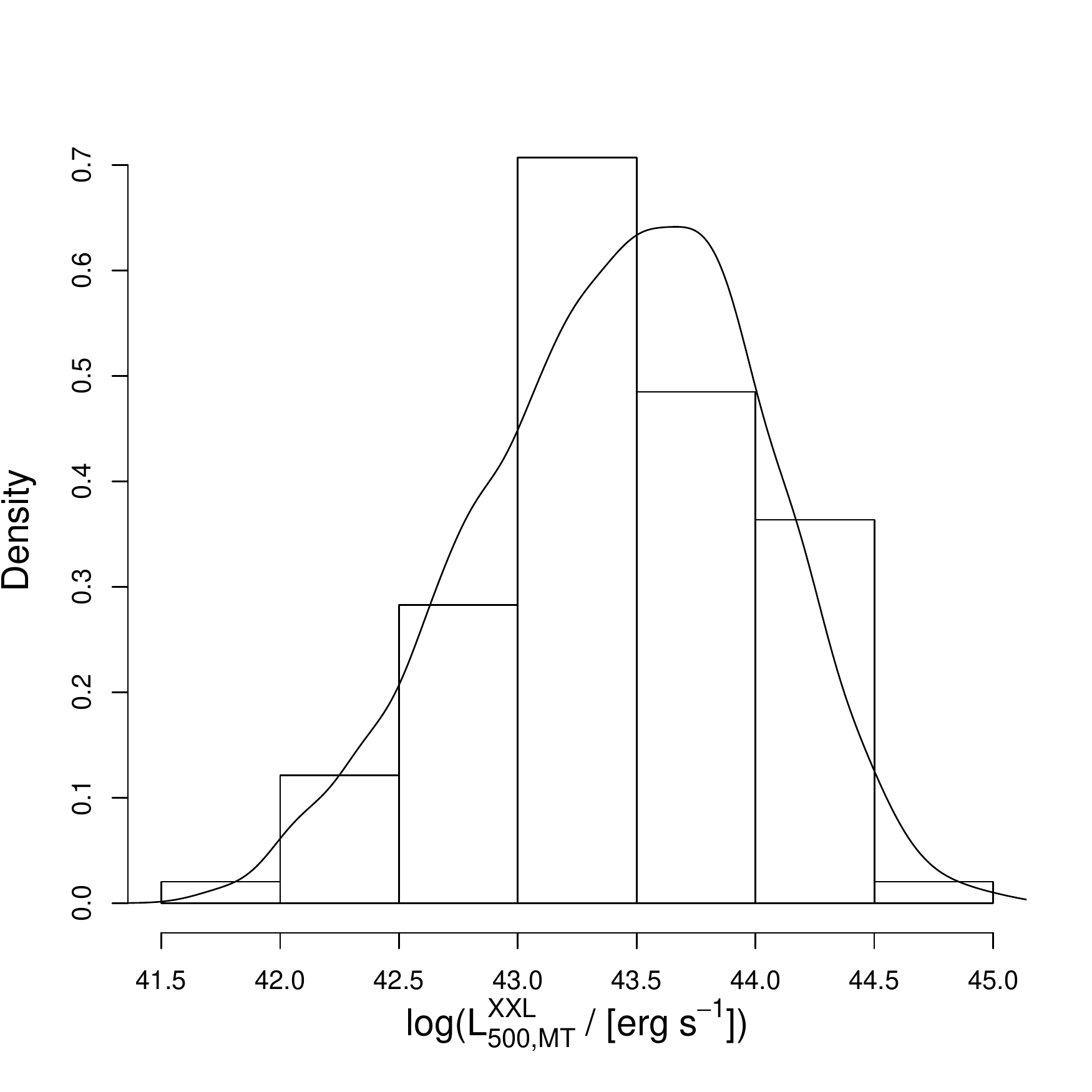}
\includegraphics[angle=0,width=6.0cm]{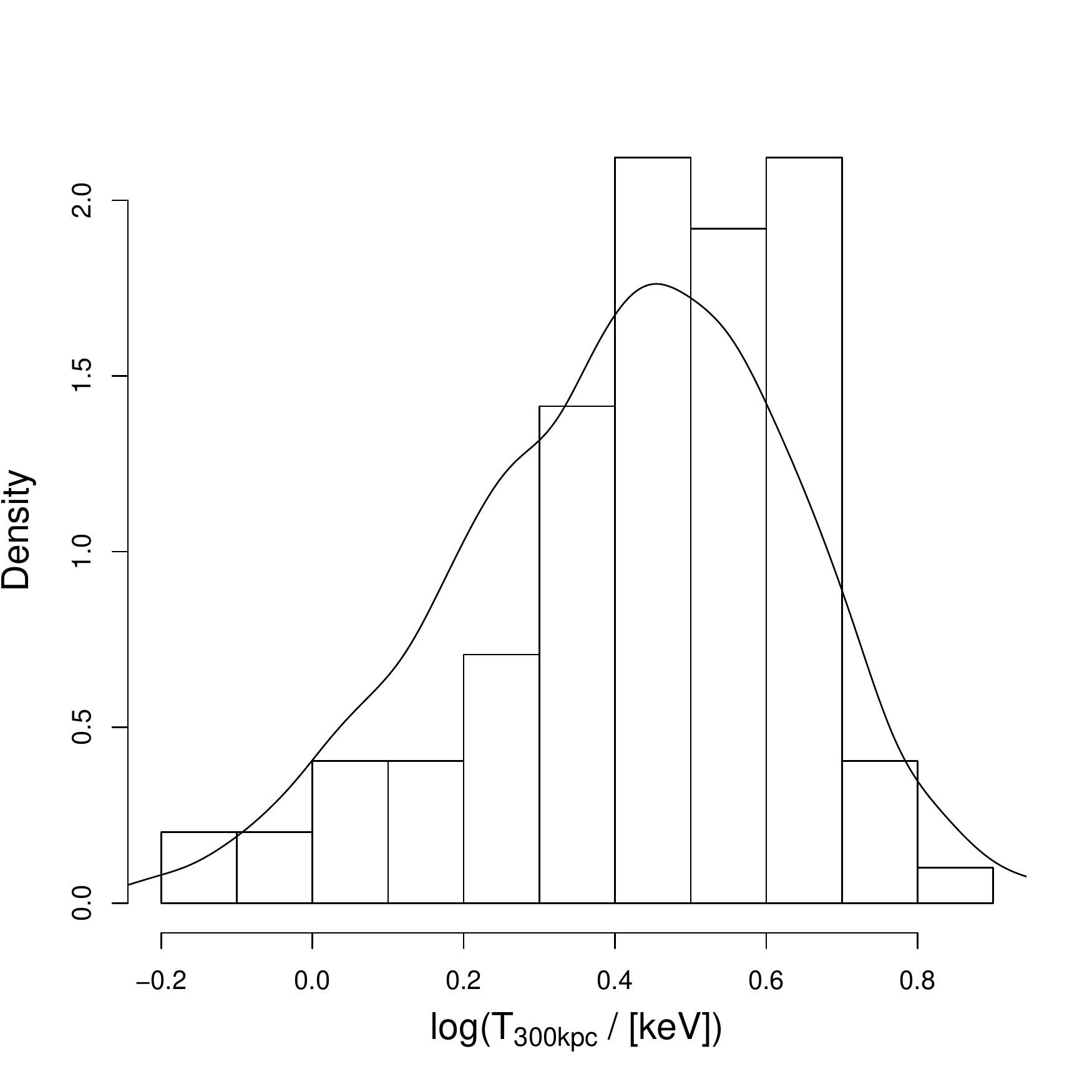}
\caption{\small{Histograms of the redshift (left panel), luminosity
    (middle panel), and temperature (right panel) of the observed
    clusters are plotted along with the density distribution of the
    same properties from the simulated clusters generated from the
    best-fitting model. The histograms and curves are normalised to a
    total area of 1.}\label{fig:hist}}
\end{figure*}

First, we make visual comparisons of the population of
clusters predicted by the best-fitting model and those observed. To
do this, a large number of model clusters were generated using the
best-fitting model parameters, and including the selection
function. Figure~\ref{fig:cont} shows contours of the simulated population
along with the observed clusters in different projections of the
observed properties. Figure~\ref{fig:hist}, instead, shows histograms of the
simulated and observed properties. In all cases the visual agreement
is good.

The second test was to compare the number of clusters predicted to be
observed by the best-fitting model with the number observed. The model
predicted 116.7 clusters, so the probability of observing a sample as discrepant as
or more discrepant than the 99 clusters observed is the sum of the
Poisson probabilities $P(N<99|116.7) + P(N>134|116.7) = 0.11$. The
number of observed clusters is thus reasonably consistent with the
number predicted.  We emphasise that this is not trivial, as the model
is not required to reproduce the number of observed clusters in our
likelihood.

The final test we make is to measure the posterior predictive
$p$-value \citep{meng1994}. This is done by
measuring the discrepancy of the data with the model for a set of the
model parameters taken from the posterior distribution, and doing the
same for a simulated data set generated from the same model
parameters. This is repeated for a large number of sets of parameters
sampled from the posterior, and the fraction of instances in which the
simulated points are more discrepant than the observed data is the
posterior predictive $p$-value. A small fraction indicates that the
model is a poor description of the data.

To implement this, we used the $\chi^{2}$ statistic to measure the
discrepancy between the observed or simulated data and the model
\citep[as in e.g.][]{2012A&A...546A...6A}. We scale the luminosity of
each point by $E(z)^{\gLT}$ for the value of $\gLT$ currently
considered and use the modified $\chi^{2}$ described in
\cite{1986nras.book.....P}, which includes the uncertainties in both
$L$ and $T$. For each sample of parameter values, a simulated
population of 99 clusters was generated and measurement errors were
assigned by taking the log-space errors from a cluster randomly
selected from the observed data;  the corresponding statistical
scatter was then added. This sampling of errors is justified, since we find
no dependency of the size of the errors on the measured parameters.

This procedure was repeated for 1,000 samples from the posterior; 
in $19\%$ of those iterations the simulated data were more discrepant
than the observed data. We thus conclude that there is no strong
evidence for the data to reject our model.

\subsection{The \lbol-$T$ Relation}
\label{sec:ltbol}

So far we have been considering the scaling of soft-band luminosity
with temperature, but it is often useful to refer to the bolometric
luminosity, $L_{bol}$.  We can convert the $\lxxl-T$ relation to
an $L_{bol}-T$ relation by using a k-correction.

Using {\tt XSPEC} simulations, we find this k-correction can be approximated
by
\begin{align}
\hspace{3.5cm}\frac{\lxxl}{\lbol} & = A_k T^{B_k}
\end{align}
with $A_k=0.587$ and $B_k=-0.450$. This power law approximates the
k-correction to within $\lesssim$3$\%$ over the temperature range of
our sample.  Substituting into the $LT$ relation we get the bolometric
relation:

\begin{align}
\hspace{2.3cm}\frac{\lbol}{L_{0}} & =
E(z)^{\gamma_{LT}}A_{LT}A^{-1}_{k}\left(\frac{T}{T_{0}}\right)^{B_{LT}}T^{-B_{k}}
\\
& = E(z)^{\gamma_{LT}}A_{LT,bol}T^{B_{k}}_{0}\left(\frac{T}{T_{0}}\right)^{B_{LT,bol}}
\end{align}

\begin{figure}[t]
\begin{center}
\includegraphics[width=8.3cm]{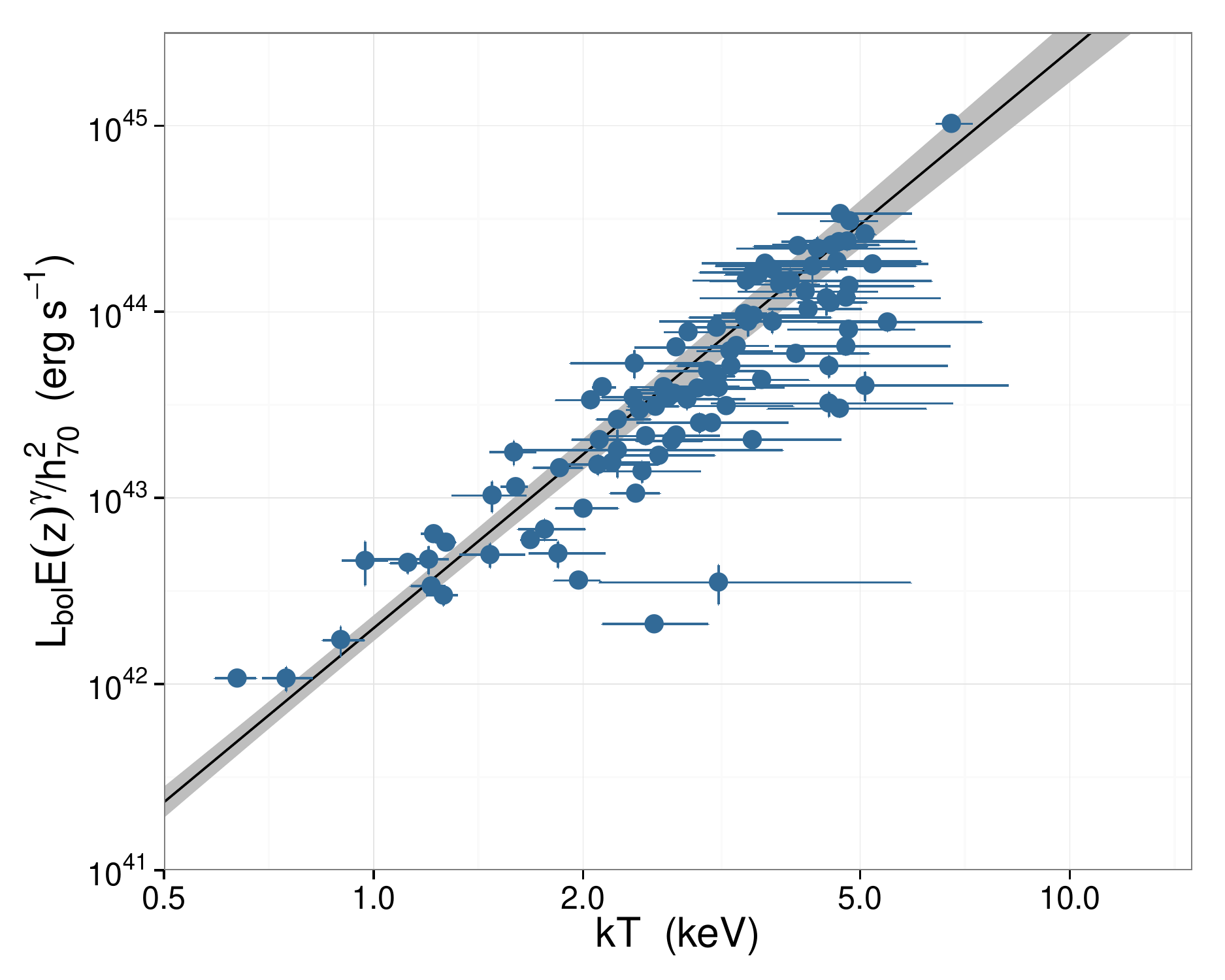}
\end{center}
\caption[]{\small{\lbol-$T$ relation with best-fitting
  model. Our best-fitting model is shown as the solid
  black line with the grey shading indicating the $1\sigma$
  uncertainty.  The best-fitting model is transformed from the soft
  band to bolometric via the steps described in
  Sect.~\ref{sec:ltbol}.}\label{fig:ltbol}}
\end{figure}

The \lbol-$T$ relation is shown in Figure~\ref{fig:ltbol};  the
best-fitting model coefficients are given in Table~\ref{tab.lt}. 

%
%______________________________________________________________

\section{Discussion}
\label{sec:disc}

In this work we have shown that the evolution of the $LT$ relation is
consistent with the strong and weak self-similar expectation
(see Sect.~\ref{sec:ltbias}).  However, our best-fit evolution is not
in agreement with previous studies, and therefore warrants further
discussion.  We will compare our results with other observational and
theoretical work and examine systematics in our analysis that could
influence the measured evolution.

\begin{table*}
\begin{center}
\begin{tabular}{lccccccc}
  \hline\hline
   Relation & Fit & Assumption & \ALT\ & \BLT\ & \gLT & \intLT &
   N$_{\rm det}$\\
   (1) & (2) & (3) & (4) & (5) & (6) & (7) & (8) \\
  \hline
  \lxxl-$T$ & XXL & $x_{500}$=0.1 & $0.63\pm0.10$ & $2.58\pm0.17$ &
  1.58$\pm$0.85 & $0.48\pm0.08$ & 117 \\
  \lxxl-$T$ & XXL & $x_{500}$=0.2 & $0.81\pm0.13$ & $2.66\pm0.15$ & 1.63$\pm$0.75 &
  $0.45\pm0.07$ & 115 \\
  \lxxl-$T$ & XXL & MT$^{\dagger}$ & $0.66\pm0.09$ & $2.47\pm0.14$ & 2.09$\pm$0.65 &
  $0.44\pm0.06$ & 178 \\
  \lxxl-$T$ & M10 & MT$^{\dagger}$ & $0.64\pm0.10$ & $2.65\pm0.15$ & 0.59$\pm$0.54 &
  $0.59\pm0.07$ & 114 \\
  \hline
\end{tabular}
\caption{\label{tab:ltcorrs} Best-fitting parameters for the $LT$
  relation while varying some assumptions in the model. (1) $LT$
  relation; (2) Fit method; (3) Assumption changed; 
   (4) Normalisation; (5) Slope; (6) Evolution term ($E(z)^{\gLT}$); (7)
  Scatter; and (8) Number of clusters predicted by the
  model. ${\dagger}$ denotes the use of the \cite{2005A&A...441..893A}
$MT$ relation.}
\end{center}
\end{table*}

\subsection{Comparison with other observational studies}
\label{sec:evol}

We found that the normalisation of the \lxxl-$T$ relation evolves as
$E(z)^{1.27\pm0.82}$ (Sect.~\ref{sec:ltbias}).
This would indicate that clusters are more luminous as a function
of increasing redshift for a given temperature.  Our result, however,
does not agree with recent studies that find a negative evolution
\citep[][C14]{2011A&A...535A...4R,2012MNRAS.424.2086H}. 

The discrepancy with C14 is surprising, given the overlap in the clusters
used from the {\em XMM}-LSS and that they account for selection
effects.  \cite{2014MNRAS.444.2723C} find a negative evolution when
using various local baseline $LT$ relations.  Assuming the
normalisation of the $LT$ relation evolves as
$E(z)(1+z)^{\alpha_{LT}}$, they find $\alpha_{LT}$=-2.5$\pm$0.4 and
$\alpha_{LT}$=-1.6$\pm$0.4 when using the P09 and
\cite{2012MNRAS.421.1583M} $LT$ relations as local baselines.  The
differences in the assumptions made in C14 and this work are the
likely cause of the discrepancy.  These assumptions are (i) the local
relation used as a reference for comparison to the high-redshift
clusters; (ii) the choice of $MT$ relation to determine $r_{500}$ and
hence $L_{500}$; C14 use the $MT$ relation of
\cite{2009ApJ...693.1142S}, whereas we use the internally calibrated
relation in Sect.~\ref{sec:analysis}; (iii) the value of $r_{c}$,
where \cite{2014MNRAS.444.2723C} use $r_{c}$=180kpc and we use
$r_{c}=0.15r_{500}$; and (iv) C14 use the likelihood from
\cite{2007MNRAS.382.1289P} for their bias correction, which does not
account for the mass function.  

To test the dependence of the fit to these assumptions, we repeated
the fit using assumptions close to those in (i) - (iii).  For this
comparison we use the $LT$ relation of P09 for a local baseline, which
gave the strongest change in evolution found in C14.  We found that we
recover a consistent, strong negative evolution on $LT$ as found in
C14 when using the same assumptions on the local $LT$ relation,
$r_{c}$, and $MT$ relation.  However, given the small dependence of
the results on the assumed $x_{500}$ (defined as the ratio of $r_{c}$
and $r_{500}$) and $MT$ relation (see
Sect~\ref{sec:assump} below), the difference in evolution is likely
driven by the assumed local $LT$ relation.  This shows that
differences in the local $LT$ relation, of less than 3$\sigma$ (see
Sect~\ref{sec:ltunbias}), can change the form of the $LT$ relation
evolution from positive to negative.  As we  discuss in
section~\ref{sec:coolcore}, this effect appears to be driven by the
differing cool core populations in the samples used for the local $LT$
relation.              

Many studies investigating the evolution of the $LT$ relation, use an
external local baseline with which to compare the normalisation of the
$LT$ relation.  This complicates the interpretation of departures from
self-similar evolution, and can change the form of the evolution (see
above).  \cite{2012MNRAS.424.2086H} investigated the evolution of the
$LT$ relation using a complete sample of 211 clusters drawn from the
{\em XMM} Cluster Survey \citep[XCS][]{2001ApJ...547..594R}.  Using an
internally calibrated $LT$ relation, they find a negatively evolving $LT$
relation of the form $(1 + z)^{-1.5\pm0.5}$. This is again in
disagreement with the positive evolution found in this work; however,
we note that the \cite{2012MNRAS.424.2086H} results do not account for 
selection biases as we have done here.

\subsection{ Impact of cool cores}
\label{sec:coolcore}

The evolving cool core population in clusters complicates the
interpretation of the evolution in the $LT$ relation. Cooling is not
present in the self-similar model, and so a reduction in the number or
strength of cool cores at high redshift \citep{2014ApJ...794...67M}
would manifest as weaker than self-similar evolution. As we saw in
Sect.~\ref{sec:evol}, using an external sample which contains a larger
number of strong cool core clusters (such as P09; see our
Fig.~\ref{fig:ltunbias}, right plot) strongly affects the magnitude
and sign of the measured evolution. It is thus possible that previous
measurements of negative evolution in the core-included $LT$ relation
 mainly reflect the decreasing contribution of cool cores to
cluster luminosities at high redshift.  An evolving mass-dependence of
the prevalence and strength of cool cores would further complicate the
interpretation. 

The  P09 sample appears to contain a population of strong
cool-core clusters, which could drive the regression fit to a higher
normalisation  not seen in the XXL-100-GC sample.  This is
likely due to a combination of the evolving cool core population and
the geometries of the two surveys.  The REXCESS clusters were selected
from REFLEX \citep{2001A&A...369..826B}, a flux-limited
(Fx[0.1-2.4 keV] $\geq$ 3$\times$10$^{-12}$ ergs s$^{-1}$ cm ${-12}$)
wide area ($\approx$14000 deg$^2$) cluster survey sensitive primarily
to high-luminosity clusters at redshifts z $\lesssim$ 0.25,   while XXL
has a much smaller local volume sensitive to lower luminosity clusters
than REFLEX.  This result was found in Paper II (see Fig. 16), where
our data was compared to REFLEX-II \citep{2013A&A...555A..30B}. 

To explain the apparent lack of strong cool core clusters in the
XXL-100-GC sample, we make the simplifying assumptions that the cool
core clusters in the P09 sample are uniformly distributed in volume
out to z=0.2 (the limit of the REXCESS sample).  The total volume of the
REXCESS sample selection was calculated from the conditions given in
Table 1 of \cite{2007A&A...469..363B}.  From the ten cool core clusters
classified in P09, we would expect an average of one cool
core cluster per 1.1$\times$10$^{8}$ Mpc$^{3}$.  When applying this to the
50 deg$^2$ area of XXL out to z = 0.2, we would expect $\approx$ 0.03
strong cool core clusters.

Thus, the evolution measured in the \lxxl-$T$ relation is not confounded
by the changing cool core population, and arguably probes the
evolution of the baryon content of the clusters more cleanly than
studies which compare high-z clusters from small area surveys with
low-redshift samples from wide area surveys.

\subsection{Systematic effects}
\label{sec:assump}

Here we investigate the dependence on the model fit to assumptions
made in this work.

\subsubsection{The choice of $x_{500}$}
\label{sec:x500}

One assumption that could have an effect on the \lxxl-$T$
relation and its evolution is the relation between $r_{c}$ and
$r_{500,MT}$ (where we adopt $x_{500}$=0.15).  To test the dependence of
the fit on the assumed $x_{500}$, we repeated the fit for
$x_{500}$=0.1 and $x_{500}$=0.2.  The results are given in
Table~\ref{tab:ltcorrs} and show  no significant difference in the
fit parameters when $x_{500}$ is varied in this range.  However,
$x_{500}$ is assumed to be independent of mass and redshift, but this
definition introduces a dependence of the physical size of $r_{c}$ on
mass and redshift, in  line with self-similar expectations.  Given
the decreasing fraction of sharply peaked cool-cores with increasing
redshift
\citep{2007hvcg.conf...48V,2011MSAIS..17...66S,2012MNRAS.420.2120M,2014ApJ...794...67M},
one could expect that the average $r_{c}$ value would increase
with redshift.  \cite{2010A&A...513A..37H} found that there was a
trend, albeit with large scatter, between $r_{c}$ and the central
cooling time (CCT) for their sample of 64 HIFLUGS clusters.  The CCT
is a robust proxy for the presence of a cool-core, and therefore this
trend indicates that $r_{c}$ increases from CC to NCC clusters
\citep[see Fig 6(b) in][]{2010A&A...513A..37H}.  Furthermore,
\cite{2010A&A...513A..37H} note a temperature dependence on $r_{c}$
such that cooler clusters appear to have smaller values of $x_{500}$
compared to hotter clusters.  Figure~\ref{fig:cont} shows the
temperature-z distribution of our clusters, highlighting how the
median cluster temperature in the sample increases with redshift.
Depending on the strength of the $x_{500}$-temperature
relation, we could be artificially introducing an evolution on
$x_{500}$.  Coupled with the $r_{c}$-CC/NCC dependance, we are most
likely underestimating the value of $x_{500}$ for the high-redshift
clusters.  This would lead us to underestimate the luminosities of the
high-redshift clusters, steepening the slope of the $LT$ relation.
This would lead to the evolution of the $LT$ relation being
underestimated.    

\subsubsection{The choice of $MT$ relation}

The choice of $MT$ relation will also have an impact on our results.
The $MT$  relation is used (i) to calculate $r_{500,MT}$ for the
cluster sample and (ii) to convert the mass function to a temperature
function (using the $M_{WL}-T$ relation, see Sect.~\ref{sec-1}).
Throughout,  we have used the $M_{WL}-T$ relation presented in Paper
IV based on XXL+COSMOS+CCCP clusters.  To test the dependence of the
fitted $LT$ relation to the choice of $MT$ relation, we repeat the fit
using the $MT$ relation of \cite{2005A&A...441..893A}, which has a
similar slope to our $M_{WL}-T$ relation but is $\approx$20\% lower in
normalisation at 3 keV.  The results are given in
Table~\ref{tab:ltcorrs}.  We find that the fit using the XXL
likelihood does not change significantly when using the
\cite{2005A&A...441..893A} $MT$ relation.  However, the fit performed
using the M10 likelihood with the
\cite{2005A&A...441..893A} $MT$ relation, would imply weaker than
self-similar evolution of the $LT$ relation.  Because of the large errors
on \gLT, the difference is only significant at the $\sim$1$\sigma$
level.  Furthermore, as a result of the overprediction of the number of
clusters in the M10 fit  (a requirement of the M10 likelihood), this fit
is not an accurate description of the data.  The predicted number of
clusters from the M10 fit using the \cite{2005A&A...441..893A} $MT$
relation, 114 clusters, does not agree with the observed number of  99
found using the Poisson calculation as above (see Sect.~\ref{sec-3}).
We note that the XXL likelihood strongly disfavours the use of the
\cite{2005A&A...441..893A} $MT$ relation in the context of the overall
model because of  the 178 predicted  clusters.  This shows that the XXL
method is less sensitive to changes in the $MT$ relation than
the M10 fitting method.  However, because of the large errors,
  drawing conclusions on the effects of the choice of $MT$ relation
  and likelihood model on \gLT\  will require many more clusters than
  contained in the XXL-100-GC sample.        

\subsubsection{The role of scatter}

Even with our efforts to model the impact of selection bias on the $LT$
relation and its evolution, we have made the simplifying assumption
that the scatter in the $LT$ relation is independent of both mass and
redshift. There is evidence to suggest that the scatter in
luminosity for a given temperature or mass decreases for higher
redshift systems \citep[][although this analysis did not account for
selection effects]{2007ApJ...668..772M}.  This can be understood from
the decline of strong cool core systems towards higher redshift
\citep[e.g.][]{2014ApJ...794...67M}.

Because the higher redshift clusters in our sample have
higher temperatures on average, the decreasing cool core fraction
 leads to a decreasing scatter in $L$ at a given $T$ for the
higher redshift part of our sample. Therefore, our model, which
assumes a constant scatter, may overestimate the scatter in the high-z
population.  This would lead to the evolution of the $\lxxl-T$
relation being underestimated since a lower scatter would require
stronger positive evolution to reproduce the observed luminous
high-redshift clusters.  

\subsubsection{The extrapolation out to $r_{500}$}
\label{sec:extrap}

Throughout this work we have used a spectral extraction region of
300 kpc for the cluster analysis.  This was chosen because
a spectral analysis within $r_{500}$ could not be achieved for the
majority of the cluster sample.  The luminosities are
determined within 300 kpc, and extrapolated out to an estimate of
$r_{500}$ by integrating under a $\beta$-profile (see
$\S$~\ref{sec:analysis}).  We investigate here what systematic effect
this could have on the derived luminosities by comparing the
extrapolated luminosities ($L^{XXL}_{500,MT}$) with those determined
via a spectral analysis within $r_{500}$ ($L^{XXL}_{500}$) for the
brightest clusters in the sample.  We also compare the temperature
derived within 300 kpc ($T_{300kpc}$) with that derived within $r_{500}$
($T_{500}$) to check for any systematic effects.

\begin{figure*}
\centering
\includegraphics[width=9.1cm]{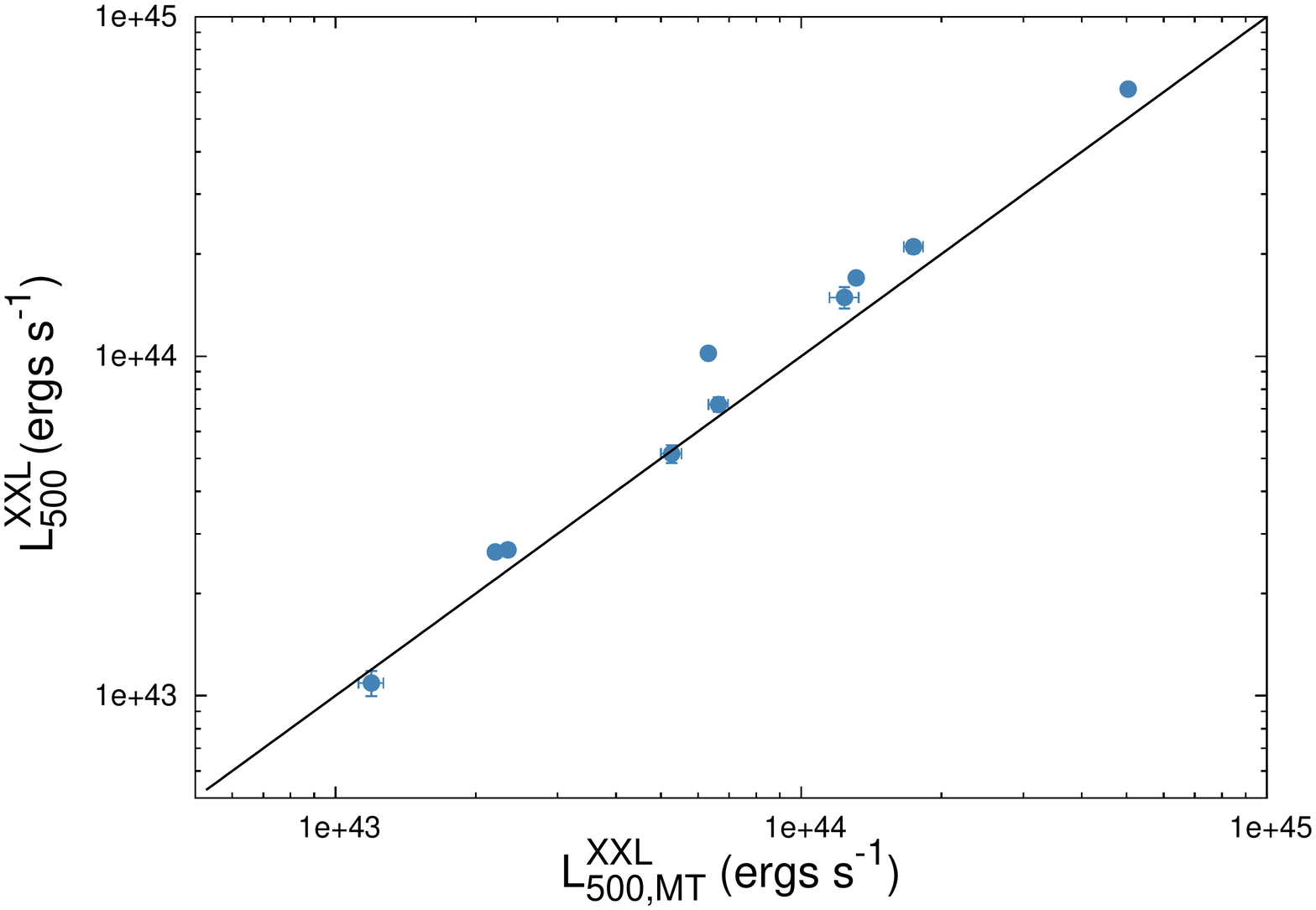}
\includegraphics[width=9.1cm]{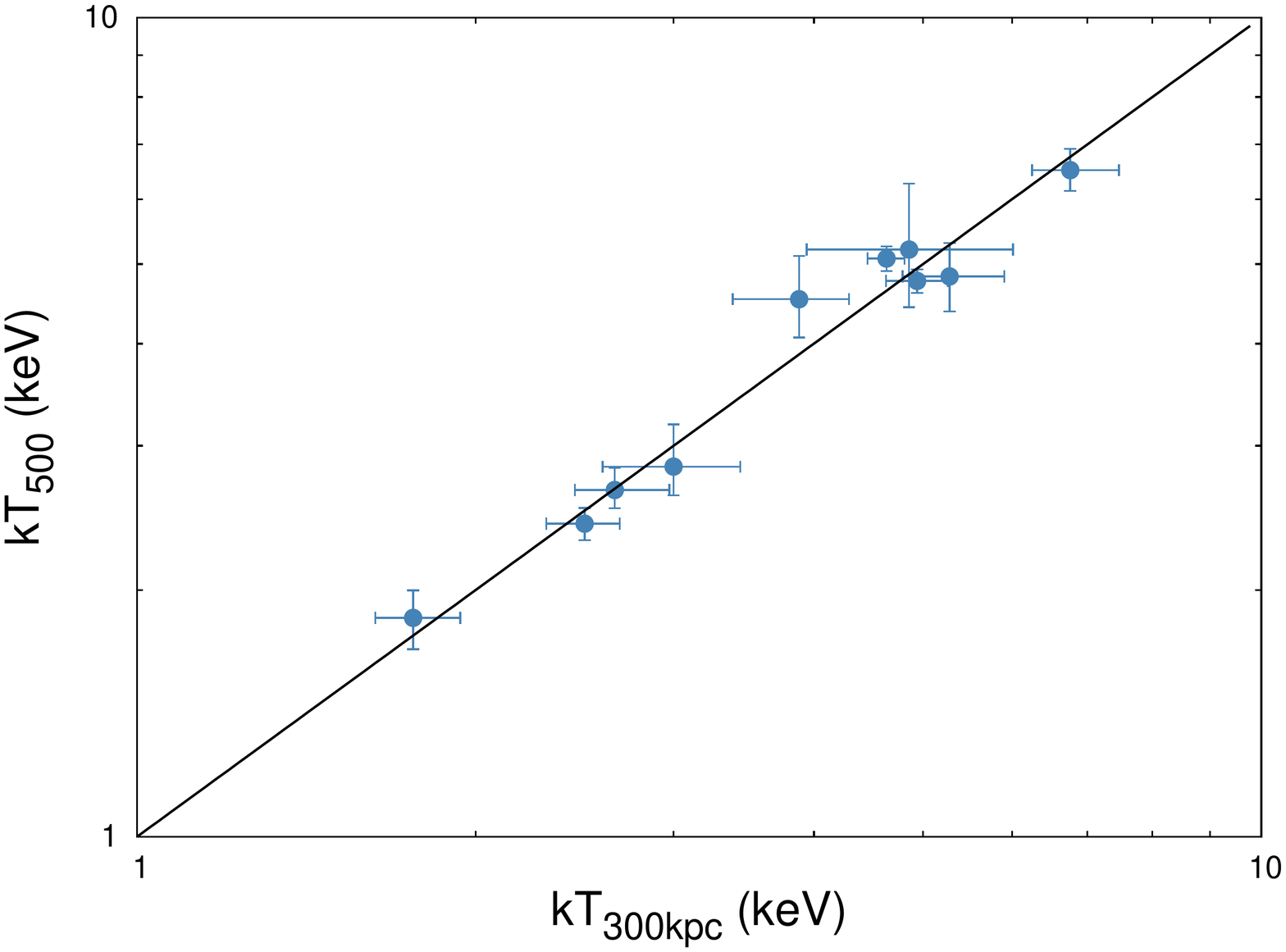}
\vspace{-1cm}
\caption[]{\small{(Left) Comparison of the luminosity extrapolated
    out to $r_{500}$ ($L^{XXL}_{500,MT}$) to that determined from a
  spectral analysis within $r_{500}$ ($L^{XXL}_{500}$). (Right) Comparison of the temperature derived within 300kpc ($T_{300kpc}$) to
  the temperature derived within $r_{500}$
  ($T_{500}$).  The black line represents a 1:1
  relation.}\label{fig:wlr500comp}} 
\end{figure*}

We estimate $r_{500}$ via an iterative process based on the $MT$
relation in Paper IV.  An initial temperature of 2 keV is used to
calculate an initial $r_{500}$, a spectrum is extracted, and a
temperature determined.  A new $r_{500}$ is calculated from this
temperature, and the process is iterated until $r_{500}$ changes by
less than 1\%.  The left plot in Figure~\ref{fig:wlr500comp} compares
$L^{XXL}_{500,MT}$ to $L^{XXL}_{500}$ for the ten highest flux
clusters in the XXL-100-GC for which the iteration process could be
performed.  We find good agreement between the luminosities for
low-luminosity clusters ($\lesssim$7$\times$10$^{43}$ erg s$^{-1}$).
However, for higher luminosity clusters ($\gtrsim$7$\times$10$^{43}$
erg s$^{-1}$) we find evidence that $L^{XXL}_{500,MT}$ could be
underestimated.  

Because of the extrapolation out to $r_{500}$, the values
of $L^{XXL}_{500,MT}$ would be underestimated by either i)
underestimating the value of $x_{500}$ and hence $r_{c}$ or ii)
overestimating the value of $\beta$.  Section~\ref{sec:x500} explored
recent evidence for a trend of $r_{c}$ with increasing temperature,
hence increasing luminosity.  The increase in $r_{c}$ with
increasing luminosity would explain the underestimate of
$L^{XXL}_{500,MT}$ compared to $L^{XXL}_{500}$, hence leading to an
underestimate of the evolution of the $LT$ relation (as
stated in Sect.~\ref{sec:x500}).  

The right plot in
Figure~\ref{fig:wlr500comp} compares $T_{300kpc}$ to  $T_{500}$.
We find no systematic differences between the two temperatures;
therefore, we can assume that $T_{300kpc}$ represents the global
temperature.  

\subsection{Do the data require $\gamma_{LT}$ as a free parameter?}
\label{sec:gammafree} 

\begin{figure}[t]
\begin{center}
\includegraphics[width=8.3cm]{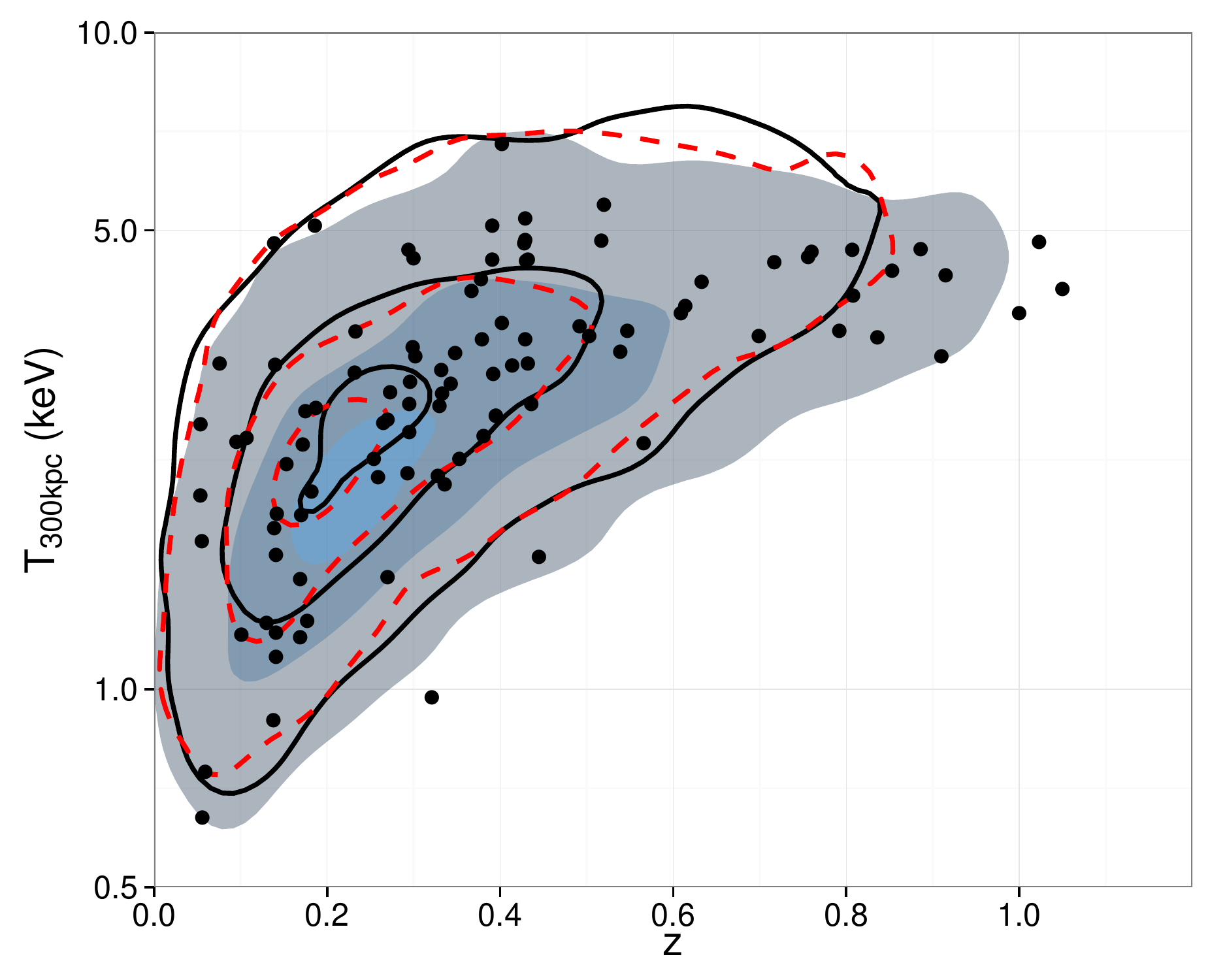}
\end{center}
\caption[]{\small{Contours of the simulated clusters generated from
    our free-evolution (shaded contours), strong self-similar (black
    solid contours), and weak self-similar (red dashed contours)
    evolution models (see $\S$\ref{sec:gammafree}).  Solid black
    points indicate the observed clusters in our
    sample.}\label{fig:ltsimcomp}}
\end{figure}

Our best-fitting evolution is consistent with strong self-similar
evolution, but only stronger than the weak self similar prediction
at the $\sim1\sigma$ level. One may question whether we are justified
in fitting for the evolution at all. In other words, compared with a
model with fixed self-similar evolution, is the improvement in the fit
when we add the additional flexibility to the model to fit the
evolution sufficiently large to justify the extra complexity in the
model? 

There are several ways to address this in a Bayesian framework, and we
adopted the deviance information criteria
\citep[DIC][]{RSSB:RSSB353,2007MNRAS.377L..74L}.  The DIC can be
computed for different models and the model with the lowest DIC is
preferred by the data with a degree of support determined by the
difference in DIC between competing models. The deviance is defined as
$D=-2\ln\lik$ and the DIC is then
given by
\begin{align}
\hspace{3.3cm}\text{DIC} = \bar{D} + p_D,
\end{align}
where $\bar{D}$ is the mean of the deviance computed over the MCMC
chain, and $p_D$ is a measure of the effective number of parameters of
the model. The latter term  penalises more complex
models, and is given by
\begin{align}
\hspace{3.3cm}p_D = \bar{D} - D(\bar{\theta}).
\end{align}
The term $D(\bar{\theta})$ is the deviance computed at the mean
parameter values from the MCMC chain.

The DIC was computed for models with $\gamma_{LT}$ as a free
parameter   fixed at the strong and weak self-similar
values of $1$ and $0.42$, respectively. In both cases, the difference  
between the free-evolution model and the fixed evolution models was
greater than -1.  Conventionally, differences in DIC more negative
than -5 are taken as strong evidence of the more complex model.  Our
analysis suggests therefore that while the data are better described
by the free-evolution model (i.e. $\bar{D}$ is lower), the improvement
is not sufficient to give strong support for including evolution as a
free parameter.

While the data as a whole do not strongly support fitting for the
evolution, we do observe that the free-evolution model appears to
provide the best description of the more distant clusters. This is
most clearly seen in the $T,z$ plane  plotted in
Figure~\ref{fig:ltsimcomp}, which compares the distribution of the
observed clusters in this plane with distributions of large
populations simulated from the best-fitting models for each
evolutionary scenario. The filled contours show the distribution of
simulated clusters from the free-evolution model; these are the same
as those shown with the contours in Figure~\ref{fig:cont} (middle
plot), but with fewer contour levels for clarity. The black solid and
red dashed contours show the distributions of clusters simulated from
the strong and weak self-similar evolution models, respectively. In all
cases the contour levels enclose $90\%$, $50\%$, and $10\%$ of the
density of simulated clusters.

Figure~\ref{fig:ltsimcomp} suggests that while the overall modelling
of the XXL-100-GC data do not strongly support fitting for the
evolution of the $LT$ relation, the fixed evolution models are not
successful at reproducing the observed population of $z>0.6$ clusters.
These clusters appear to prefer stronger than (weak) self-similar
evolution, in order that they are sufficiently luminous to be included
in the XXL-100-GC sample.  With this in mind, it is worth considering
the interpretation and implications of a stronger-than-self-similar
evolution in the $LT$ relation.

\subsection{Interpreting $\gLT$}
\label{sec:intglt}

The evolution of the $LT$ relation is best expressed as a
combination of the evolution and slopes of the $LM$ and $MT$ relations
\citep[e.g.][]{2014MNRAS.437.1171M},
\begin{align}
\hspace{2.5cm}\gLT & = \gLM - \frac{B_{LM}}{B_{TM}}\gTM,
\end{align}
where $B_{LM}$ and $\gamma_{LM}$ are the slope and power of the
evolution term of the $LM$ relation, respectively. We note that $B_{TM}$
is the reciprocal of the slope of the $MT$ relation, $B_{MT}$.

Thus, stronger-than-self-similar evolution in the $LT$ relation
implies stronger-than-self-similar evolution in the $LM$ relation, or
weaker-than-self-similar evolution in the $TM$
relation. Stronger-than-self-similar evolution in the $LM$ relation
requires either that the baryon fraction within a fixed overdensity
be higher at high-z \citep[disfavoured by simulations,
e.g.][]{2013MNRAS.431.1487P,2013ApJ...776...81B} or that it  be driven by
higher density regions of ICM at high-z. In the absence of cool cores,
such regions could perhaps be associated with the larger degree of
substructure typically seen in higher redshift clusters as we approach
the epoch of cluster assembly
\citep[e.g.][]{2007ApJ...658..865J,2008ApJS..174..117M}.

Alternatively, weaker-than-self-similar evolution of the $TM$ relation
would require clusters to be cooler for a given mass at
high redshift. This could be caused by incomplete virialisation of the
ICM at higher redshift. Given that our analysis assumes self-similar
evolution of the $TM$ relation, a weaker evolution would mean that we
underestimate the mass and hence $r_{500}$ for higher redshift clusters in
our Bayesian analysis. This in turn would mean that the luminosity
extrapolated to $r_{500,MT}$ would be underestimated and that the core radius
(defined via $x_{500}$) would be underestimated. An underestimation of the
core radius for high-z clusters would generally mean that we
overestimate the detection probability for those clusters. This would
lead us to underestimate the evolution in $LT$, since fainter clusters
would have a higher detection probability than if the core radius were
larger. These arguments together imply that any inference of
weaker-than-self-similar (i.e. more negative) evolution of the $TM$
relation could be regarded as an upper limit, in the sense that the true
evolution could be weaker still.

\subsection{Comparisons with simulations}      
\label{sec:sims}

Since we find a positive evolution of the $\lxxl-T$ relation, we compare our data to
simulations in order to determine what physical processes could be at
play that give rise to our observed evolution.
\cite{2010MNRAS.408.2213S} studied the evolution of scaling relations
using simulations from the Millennium Gas Project
\citep{2005Natur.435..629S}.  They employ three different models in
their simulations, a gravity-only (GO) control model, a simple
preheating model (PC), and a model using feedback from the energy
input due to SNe and AGN (FO). Firstly, they show that the local $LT$
relation for the PC and FO runs compare well to the observed $LT$
relation, finding $B_{LT,S10}\approx3$.  Secondly, they find that the
evolution of the $LT$ relation behaves differently for the PC and FO
simulations.  The $LT$ relation evolves negatively for the PC simulation,
whereas the FO simulation leads to a positively evolving $LT$ relation.
The evolution found using the FO simulation follows the positive
evolution of the \lxxl-$T$ relation found in this work.  

As shown in Section~\ref{sec:intglt}, the interpretation of the
evolution of the $LT$ relation is complicated owing to the dependence of
the evolution on the $LM$ and $TM$ relations.
\cite{2010MNRAS.408.2213S} found that the evolution of the $TM$
relation evolves negatively, a result implied by the observed
positive evolution of the \lxxl-$T$ relation.  Negative evolution of
the $TM$ relation was also found in \cite{2014MNRAS.445.1774P}, using
simulations taking into account radiative cooling, star formation,
supernovae feedback, and AGN feedback.  This provides evidence that the
positive evolution observed in the \lxxl-$T$ relation is
underestimated (see Sect.~\ref{sec:intglt}).  Furthermore,
\cite{2010MNRAS.408.2213S} and \cite{2014MNRAS.445.1774P} find a
positively evolving $LM$ relation with a slope steeper than the
self-similar expectation.  A positively evolving $LM$
relation with a steeper-than-self-similar slope and negatively
evolving $TM$ relation (the slope of the $TM$ relation from the
simulations agree with the self-similar expectation) would lead to a
positively evolving $LT$ relation.  Therefore, the evidence of a
positively evolving \lxxl-$T$ relation (see Sect.~\ref{sec:ltbias}) is
in line with the expectation from simulations.  

%
%______________________________________________________________

\section{Conclusions}
\label{sec:conc}

We have presented a detailed analysis of the brightest 100 clusters
detected in the 50 deg$^{2}$ {\em XMM}-XXL Survey;  the sample spans a
wide range of redshift (0.05 $<$ z $<$ 1.05), temperature (0.6 $<$ $T$
$<$ 7.0 keV), and luminosity (9$\times$10$^{41}$ $<$ $L$ $<$
5$\times10^{43}$ erg s$^{-1}$).  The $LT$ relation has been studied in
detail and we present the first measurement of the evolution using a
single sample with an internal local baseline $LT$ relation and fully
accounting for selection biases.  
 Our main conclusions are as follows.
   \begin{enumerate}
      \item The sample $\lxxl-T$ relation has a slope of
        $B_{LT}$=3.01$\pm$0.27, not considering the effects of
        selection biases.  This is consistent with previous studies of
        the observed steep slope of the $LT$ relation when compared to
        the self-similar expectation.
      \item When taking into account the selection effects utilising a
        Bayesian approach using two forms of the likelihood we find
        a slope of the soft-band $\lxxl-T$ relation of
        $B_{LT}$=2.63$\pm0.15$ and a bolometric slope of
        $B_{LT}$=3.08$\pm0.15$.   
      \item  After taking into account the
        selection effects, our data show that the $\lxxl-T$ relation
        prefers an evolution of the form $E(z)^{1.64\pm0.77}$.  This is
        consistent with the expected ``strong'' self-similar
        evolution; however, is marginally stronger than the ``weak''
        self-similar expectation.  
      \item Comparisons to clusters detected in the {\em XMM}-LSS (a
        precursor to XXL), which favour a negatively evolving $LT$
        relation, can be explained by the different assumptions made
        for the local baseline $LT$ relation, the assumed mass-temperature
        relation, and the value of the core radius.
        \item Investigating the impact of the assumed
          mass-temperature relation and the $x_{500}$ parameter, we
          find that they do not have a significant impact on the $\lxxl-T$
          relation when using the XXL likelihood model.  However,
            because of the large errors on the evolution, there is a
            degeneracy between the choice of MT relation, likelihood
            model, and evolution.
        \item We find that small changes in the comparative local
          baseline $LT$ relation can change the inferred evolution of
          the $\lxxl-T$ relation.  This appears to be due to the
          differing cool core populations in the samples used for the
          local $LT$ baseline, which are affected by the evolving cool
          core population and the geometries of the surveys used to draw
          the cluster samples used for the $LT$ baselines.
        \item By comparing our results with those determined from
          simulations, we find that the positive evolution favours
          models of cluster formation that include feedback from
          energy injection from SNe and AGNs.   
   \end{enumerate}

Our results show that the evolution of the $LT$ relation is strongly
affected by the  choice of comparative local baseline scaling
relations.  The {\em XMM}-XXL Survey has allowed us for the first time
to study in detail the evolution of the $LT$ relation, fully accounting
for selection biases, using a single homogeneous sample of clusters.
Furthermore, using the data from the $\approx$450 clusters detected in
the XXL survey, we will be able to place robust constraints on
cosmological parameters.                

\begin{acknowledgements}
XXL is an international project based around an XMM Very Large
Programme surveying two 25 deg$^{2}$ extragalactic fields at a depth
of $\sim$ 5$\times$10$^{-15}$ erg s$^{-1}$ cm$^{-2}$ in the [0.5-2.0]
keV band for point-like sources.  The XXL website is
http://irfu.cea.fr/xxl. Multiwavelength information and spectroscopic
follow-up of the X-ray sources are obtained through a number of survey
programmes, summarised at http://xxlmultiwave.pbworks.com/.  PAG, BJM,
ML, GPS, TP and JD acknowledge support from the UK Science
and Technology Facilities Council.  GPS acknowledges support from the
Royal Society. FP acknowledges support from the BMBF/DLR grant 50 OR 1117, 
the DFG grant RE 1462-6 and the DFG Transregio Programme TR33.
\end{acknowledgements}

%-------------------------------------------------------------------

\bibliographystyle{aa}
\bibliography{xxl_lt.bib}

\appendix

\end{document}